\documentclass[12pt,letterpaper]{article}
\usepackage{amsmath}
\usepackage{verbatim}
\usepackage{setspace}
\usepackage{graphicx, xcolor, color}
\usepackage{url}
\usepackage{amssymb}
\usepackage[sectionbib]{natbib}
\usepackage{sectsty}
\usepackage{multirow}
\usepackage{longtable}
\usepackage{rotating}
\usepackage[labelfont=bf,font=footnotesize]{caption}
\usepackage[left=2.56cm,right=2.56cm,top=2.56cm]{geometry}
\usepackage{mathrsfs}
\usepackage{dsfont} 
\newcommand{\bm}[1]{\boldsymbol{#1}} 
\newcommand{\R}{\textsf{R}\space} 

\def\citeapos#1{\citeauthor{#1}'s (\citeyear{#1})}

\begin{document}

\title{What can we Learn from Predictive Modeling?\footnote{Many thanks to Alison Craig for research assistance. Sincere thanks also to Matt Blackwell and Michael Neblo for helpful comments on an earlier draft. The authors are grateful for the support of the National Science Foundation (SES-1558661, SES-1619644, SES-1637089, CISE-1320219, SES-1357622, SES-1514750, and SES-1461493) and the Alexander von Humboldt Foundation. Replication data are posted to the {\em Political Analysis} Dataverse \citep{replication_data}.}}
\author{Skyler J. Cranmer \and Bruce A. Desmarais\footnote{Skyler Cranmer is the Carter Phillips and Sue Henry Associate Professor in the Department of Political Science at the Ohio State University (\texttt{cranmer.12@osu.edu}{cranmer.12@osu.edu}). Bruce Desmarais is an Associate Professor in the Department of Political Science at Pennsylvania State University (\texttt{bdesmarais@psu.edu}{bdesmarais@psu.edu}).}}
\maketitle

\begin{abstract}
\noindent The large majority of inferences drawn in empirical political research follow from model-based associations (e.g. regression). Here, we articulate the benefits of predictive modeling as a complement to this approach. Predictive models aim to specify a probabilistic model that provides a good fit to testing data that were not used to estimate the model's parameters. Our goals are threefold. First, we review the central benefits of this under-utilized approach from a perspective uncommon in the existing literature: we focus on how predictive modeling can be used to complement and augment standard associational analyses. Second, we advance the state of the literature by laying out a simple set of benchmark predictive criteria. Third, we illustrate our approach through a detailed application to the prediction of interstate conflict. 
\\~\\ \begin{center} {\large  Forthcoming in {\em Political Analysis}} \end{center}
\end{abstract}

\doublespacing

\vspace{1cm}

\begin{center}
\emph{Word count: 9,205}
\end{center}

\thispagestyle{empty}
\setcounter{page}{0}
\newpage

\section{Introduction}





Most empirical political science research relies on model-based associations (e.g., regression) in observational data to test hypotheses and develop explanations of the phenomena under study \citep{druckman2006}. 
Much emphasis is typically placed on the theoretical specification of a statistical model, which we agree is important, while much less emphasis is placed on evaluating the predictive performance of the model. 
We propose thinking about the role of prediction in theory-building as a continuum, in which standard models are subject to increasingly strong predictive tests. 
Key distinctions occur where prediction-based validation is shifted from in-sample (e.g. predicting the data used to fit the model) to out-of-sample (e.g. predicting data not used to fit the model) and again when prediction is used to learn about the process of interest independent of existing theory rather than to validate a theoretically driven model. 
In the current article, we (1) make the case that predictive models are under-used in political science, (2) elucidate what we see as their most attractive features, and (3) demonstrate how prediction can augment association-based modeling and even lead to new discoveries. 

The chronic lack of emphasis on model validation in political science risks a situation in which most  inferences rely on models that might fit poorly and makes the contributions of new research on established topics ambiguous at best. 
What is more, the field's reticence to use prediction often prevents us from refining our measures and models, and making objective comparisons of the performance of competing theories. 
Science is meant to be a cumulative enterprise, but the lack of clear, performance based, model evaluation makes it difficult, if not impossible, to judge the relative contribution of new empirical work relative to the existing literature. Furthermore, the lack of cumulative/benchmark predictive assessments renders it similarly difficult to judge overall scientific progress on a specific outcome. All of this together serves to limit on our ability to advance political inquiry.

We have three goals in this work. 
First, we catalogue the dangers associated with conducting model based inference without assessing predictive performance, while also pointing out some ways in which assessing predictive performance can bolster our inferences. 
Second, we lay out the criteria for what we believe should constitute a benchmark predictive model: one based either on the ``state of the literature'' model if one is available or on the structure endogenous to the outcome variable if the researcher is establishing his or her own baseline, both using out-of-sample predictive accuracy as the sole criterion for model quality. 
Third, we illustrate our approach in application to the prediction of the initiation of serious inter-state conflicts. This illustration yields several interesting, and perhaps unexpected, results: models based only on the endogenous structure of the outcome variable generally outperform models based on exogenous predictors, models combining endogenous and exogenous predictors generally predict worse than models based only on the outcome variable, and many of the best established variables in the literature contribute little to the model's predictive accuracy. 
In summary, we aim to explicate and illustrate how the evaluation of predictive performance can be better utilized with the end goal of strengthening the model based inferences on which we so often rely to advance the state of knowledge in our field.

\section{Prediction and Inference}

The technical distinction between inferential modeling and predictive modeling is rather straightforward, though practical distinction for the applied researcher is much less so.  In inferential modeling, the statistical model is constructed as an operationalization of a theoretical model. The specification is important because deviations from the theoretical model in operationalization inhibit our ability to use the statistical model to test hypotheses. The coefficients are the objects of interest, which is to say that the \emph{statistical model itself is the object of interest}. In inferential modeling, we use the data to learn about the statistical model. Conversely, in pure predictive modeling, the objects of interest are the variables rather than the parameters: we use the available data to produce the best possible predictions of the outcome variable. It does not matter, for purely predictive exercises, whether the statistical model used is a close operationalization of a causal theory, because the only metric for the quality of a model here is its predictive performance \citep{Shmueli:2010}.  A subtler difference is that inferential models aim to minimize bias in order to produce the most accurate coefficient estimates, whereas predictive models minimize the combination of bias and estimation variance in order to optimize empirical precision \citep[p. 293]{Shmueli:2010}. 

\begin{figure}
\centering
\includegraphics[width=\textwidth]{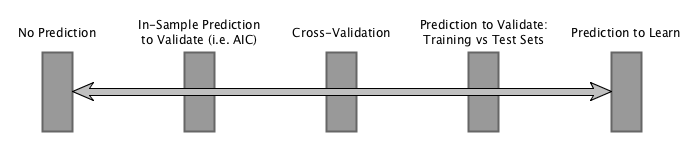}
\caption{A conceptual continuum between no prediction and pure prediction. Political science research tends to fall around the left side of this continuum whereas fields like computer science tend to fall to the right of it.}
\label{continuum}
\end{figure}

When we say that the practical distinction between inference and prediction is less straightforward, we mean that inference can be augmented by prediction. 
Possibly greatly. 
Consider a continuum in which prediction is applied increasingly strongly.  
This continuum is illustrated in Figure \ref{continuum}. 
On one end of the continuum, the researcher does not use predictive methodology at all. Here, no validation of the model takes place, the researcher simply runs the model, interprets the results, and concludes. The following section will make clear that this approach suffers from a number of problems that could be ameliorated by predictive modeling. 

Moving right on the continuum, we consider in-sample approaches to prediction. ``In-sample'' means that the same data used to fit the inferential model are used in the predictive exercise. Examples of in-sample techniques that are frequently applied include the $R^2$ and AIC statistics. Some other common techniques, such as plotting the receiver operating characteristic (ROC) curve \citep{fawcett2006} or posterior predictive checks in a Bayesian context \citep{Gill:2014}, can be applied either in-sample or out-of-sample.

As we move further right on the continuum, we reach the first of two conceptually important cutpoints: the point at which validation-driven prediction is made out-of-sample as opposed to the in-sample prediction further left on the continuum. Here we must draw a distinction between two samples (datasets) used in predictive modeling: the training set and the test set. The training set is the set of data upon which the predictive model is built. One generally tries to capture the process of interest in the training set, often by making iterative adjustments to the statistical model. In out-of-sample prediction, the model produced on the training set is then applied to the test set to estimate {\em generalization error}. Generalization error is the prediction error of a model when applied to the general population of interest (i.e., beyond the sample on which the model was trained) \citep{nadeau2003}. In-sample-prediction, by contrast, functions similarly on the training set, but then tests predictive accuracy on the training set as well (a test set is not used in in-sample-prediction) \citep{10.1525/j.ctt13x1gcg.8}.

Our search and review of the literature suggests that the majority of political science analyses fall into the left two categories of this continuum, using in-sample validation or no validation at all. To provide some evidence for this claim, we searched JSTOR (dated 5-16-2016) for (``cross validation'' OR ``out of sample'') published in political science journals since 2005, and found 283 results. As a point of comparison, a search for (``logistic regression'' OR ``logit'') with the same search parameters returned 3,151 results. Our aim for the remainder of the article is to convince the reader that our field can profit from occupying spaces of the continuum further right.

Cross-validation can take many specific forms, but generally involves randomly dividing the data into several partitions, fitting the model of interest on all data not in a given partition, and then testing the the model with the held-out partition. This process is then repeated for all partitions and the mean error is reported \citep{Stone:74, Stone:1977}. Depending on the size of the dataset and the number of test partitions used, this technique may be computationally expensive \citep{faraway2006extending}. Cross-validation has been used in every subfield of political science. For example \cite{Beck:2000} use it to evaluate the performance of several conflict models they had fit.

A simple and powerful alternative to cross-validation is to hold the training and test sets as completely distinct datasets. Typically, a test set of 30-50\% of the primary data is randomly sampled and set aside while the model is trained on the remainder \citep{10.1525/j.ctt13x1gcg.8} This setup has the elegant feature that, because the test set was randomly partitioned from the training set, the only thing the two have in common in expectation is the data generating process. Thus, if a model fits the test set well, one can expect that key elements of the data generating process are captured in the theoretically informed model. For example, \cite{Goldstone:2010} found that a relatively simple model greatly increases predictive power and casts doubt on the role of established covariates in the prediction of civil conflicts. More recently, applying this concept to longitudinal data \cite{Cranmer2015} find that inclusion of their ``Kantian Fractionalization'' measure---a summary measure of degree of clustering and cluster cohesiveness across international networks of trade, IGO, and joint democracy---adds more to the predictive performance of a typical conflict model than all the standard control variables combined and that joint democracy makes a negligible contribution to the prediction of conflict.  

The rightmost extreme of the continuum uses the training/test set setup to learn, as opposed to prediction to validate, in an approach often called \emph{machine learning}. This is the second conceptually important cutpoint on the continuum: in this extreme space of the continuum, one is no longer seeking to validate a theoretically informed model, but seeking to learn a model from the data by minimizing generalization error through predictive experiments.  A hallmark of machine learning is that the training set may be mined in an unsupervised manner (e.g. with no specific model specified and no application-specific rules as to how the mining should be conducted) with the goal of finding a specification that predicts the test set as well as possible. In other words, machine learning algorithms are designed to perform better with more data; they ``learn'' from the the data they have experienced and they learn more from more experience. As \citet[p. 978]{hua2009} note, concisely, ``Machine learning is a subject that studies how to use computers to simulate human learning activities.'' 

Machine learning can be useful for theory building because it can uncover patterns that might not have been obvious or intuitive to the theory-building analyst, and, on the other hand, can suggest features or sets thereof that should be excluded from the model altogether. However, the machine learning approach is relatively uncommon in political science. Consider the example of \cite{Desmarais:2011ieee}: considering a large set of measures computed endogenously on the network of transnational terrorist attacks (an individual from country $i$ attacks a target in country $j$), \cite{Desmarais:2011ieee} mine the set of specifications to produce a model that predicts new attacks out of sample with more than 95\% accuracy and with probabilities assigned to attacks that ultimately occur several orders of magnitude higher than attacks that ultimately do not occur. More recently, \cite{Muchlinski:2016} found that a machine learning approach significantly increases the predictive accuracy of civil-war models.



\section{The Utility of Prediction}

Understanding the statistical differences between prediction and explanation is necessary to elucidate the distinctive utility of the two endeavors. More important however are the contributions predictive modeling can make to our explanatory understanding. The contributions are many, leading us to claim that an exclusive focus on explanatory modeling omits a great deal of leverage predictive modeling can lend to the explanatory exercise.

\subsection{Systematically Observing Nature}

In theory, most political science research begins with a novel hypothesis, and follows the model of hypothetico-deductivism in which empirical expectations are deduced from the hypothesis, with empirical tests to follow \citep{clarke2007}. 
This skips a crucial step in the scientific process: exploratory observation of nature. To be clear, nature here refers to the political processes of interest, and also the environments in which they occur. Observation is critical to forming new hypotheses because it is the observation of associations and the consideration of their potential causal relationships that forms the backbone of theoretical development. Yet typically we rely on our reading of history and the existing literature to constitute our observation of nature. The use of predictive models can uncover unknown patterns and new causal mechanisms in complex data. The first thing predictive modeling offers us is the opportunity to observe nature in a systematic way. By finding new patters, inductively, we may form new hypotheses about why those patterns exist and, through subsequent tests, improve the state of our science \citep{Gurbaxani:1990, Gurbaxani:1994, Collopy:1994, Shmueli:2010}.

Not systematically observing the phenomena of interest prior to hypothesis formation involves the rather bold claim that we do not need such empirical tools because our powers of observation are so keen that we are able to detect \emph{all} meaningful patterns in the extremely complex phenomena we study, so as to be able to completely and correctly specify not only our theories, but our explanatory statistical models (where complete and correct specification is  a statistical necessity if one hopes to test a theory). 
Particularly as political science moves into the era of ``big data,'' it seems increasingly unlikely that a researcher will be able to detect all meaningful patterns without a predictive model.


\subsection{Refining Measures and Models}

Predictive models can play an important role in the refinement of both measures and explanatory statistical models. In the following list we denote three related ways in which predictive assessment can aid in improving explanatory models.

\emph{Refining Measures}. Predictive modeling can help us refine our operationalizations of important theoretical concepts. This can be accomplished in two ways: predictive exercises may be conducted to discover new measures \citep{VanMaanen:2007, Shmueli:2010} or to test the efficacy of competing operationalizations against one another \citep{Shmueli:2010}. Equally, if not more usefully, predictive accuracy is an impartial criterion by which to evaluate competing operationalizations of the same concept.

\emph{Refining Models 1: Parsimony.} Predictive models provide a means for impartially evaluating the parsimony of an explanatory model \citep{Jensen:2000}. There has been some debate in the literature about how parsimonious an explanatory model should be, with \citet{Achen:2002} arguing for few (3) variables, others arguing that more is better to avoid omitted variable bias \citep{Oneal:2005}, and many more moderate perspectives in between. Yet prediction, especially out-of-sample prediction, is a useful way to tune the parsimony of a model because predictive exercises allow one to judge how much impact is being made by each element of the model in terms of its contribution to predictive performance \citep{Hastie:2009, Kuhn:2013}. 

\emph{Refining Models 2: Diagnosing Misspecification without Overfitting.} Out-of-sample predictive exercises can be used to identify model misspecification without running the risk of over-fitting the data. Predictive models are natural tools for identifying model misspecification because misspecified models are necessarily poor predictors. However, overfitting -- including excess parameters that exploit artifacts of the data without capturing the data generating process -- can often disguise misspecification by moving the model towards saturation. When predicting out of sample, the latter disguise will not work and the prior problem will be apparent.

\subsection{Objective Comparison of Model Quality\\ and Competing Theories}

Predictive models afford the researcher the ability to test the quality of an explanatory model against more realistic null models or even against rival theoretical models. 

\emph{Better Null Models.} Truly null models, where there is absolutely no relationship between a dependent variable and a set of independent variables, are quite rare in political science. Why then do we test our theoretically informed models against null models that are so unlikely to occur? Thinking about it this way, the claim that ``my model fits the data better than a spectacularly unlikely model'' rather takes the zip out of claiming statistical significance for an effect. As we will explain below, a baseline comparison model (even a naive one) need not be so simple as a null model. Rather, one can set reasonable, if simple, criteria for benchmark models and use predictive accuracy as a means by which to judge if one's model outperforms the baseline. 

\emph{Testing Competing Theories.} Predictive ability as an excellent way to compare competing theories of the same outcome. Using predictive criteria, particularly out of sample predictive criteria, is an exceedingly simple means to highlight the extent to which the theoretically informed models anticipate reality, and which among those models does a better job of it. Measures of in-sample fit (e.g. adjusted-$R^2$, AIC, BIC), and even in-sample prediction, are less-than-ideal because they run the risk of overfitting the data; accidentally exploiting artifacts of the error term that are not part of the data generating process in nature. Perhaps more importantly, certain elements of the competing theories may be mutually exclusive or highly collinear, making comparative testing without relying on predictive criteria all but impossible. While direct comparisons of model fit with non-nested sets of predictors is possible in a Bayesian context, they are not in a frequentist context \citep{Gill:2014}. Yet, using out-of-sample predictive fit to judge one model against competing specifications (even if they are very similar) is straightforward in either a Bayesian or frequentist approach and the improvement of model one over the other can be measured precisely and objectively. For example, in the study of environmental politics it is well established that both a county's population and GDP affect its CO$_2$ emissions. But when including CO$_2$ emissions as a predictor, should this measure be normalized by population or GDP? A predictive approach, even one as simple as cross-validation, could do much to disentangle these closely related but theoretically distinct predictors. 

\subsection{Measuring the State of Knowledge}

Finally, predictive modeling can tell us how well an explanatory theory captures the phenomena of interest and can provide an upper bound on what can be learned by further explanatory modeling. Suppose we examine the dominant theory of a particular outcome, say interstate war, and find that its predictive accuracy is quite low. This tells us that one of two things, or both together, is happening: either the dominant causes of the outcome have yet to be discovered and much important work remains to be done in the field, or the outcome exhibits a high amount of ``noise'' and a comparatively small amount of natural ``signal.'' In the first case, we may conclude that our theoretical understanding of the phenomena is grossly incomplete and use this conclusion to fuel a push for improved theory and causal testing. In the second case, we may have arrived at a largely complete explanatory model of the phenomena, but high degrees of imprecision in our ability to measure the relevant variables produce what appears to be high stochasticity and prevents accurate predictions. As \citet[p.4]{Shmueli:2010} notes, ``Predictive modeling plays an important role in
quantifying the level of predictability of measurable phenomena by creating benchmarks of predictive accuracy.'' 

Consider alternatively what the field would learn were the dominant theory to produce a high level of predictive accuracy. If the state of the art predicts well, that tells us that there is comparatively little that may be gained from continued efforts at explanatory modeling. If the addition of theory can, at most, produce a marginal increase in predictive accuracy, then such theory is only capable of giving us marginally more traction on the problem. 


\section{Establishing a Predictive Baseline}




Above, we considered the utility of having a simple but non-null benchmark model against which to compare the predictive power of theoretically informed models. Such benchmark models can take two forms: they can either reflect the most recent or best-accepted model already established in what we will call a ``state-of-the-literature'' model, or they can reflect the best model one can specify without relying on theory in what we call a ``baseline'' model. In principal, using a state-of-the-literature model is straightforward. In practice however, such models can be difficult because (a) most political science research does not provide predictive results against which an analyst may easily compare new results and (b) replicating the state-of-the-literature model to produce such a predictive benchmark is often not as easy as it should be.\footnote{The latter of these problems is, hopefully, decreasing over time as it becomes more common for researchers to publish their replication data/code and an increasing number of journals are requiring such replication materials to be made public.} In situations where a state-of-the-literature model is not available or not desired, establishing a reasonable baseline model will be important. Indeed, one may wish to establish a baseline model even when a state-of-the-literature model is available in order to judge the predictive gains offered by the state-of-the-literature model. 

Here, we propose general criteria for baseline benchmark models. The proposed criterion, we argue, is the best model one can specify without reflecting the proposed theory. As such, the baseline model is similar to null model comparisons common in statistical mechanics, but less naive.  Such baseline models are useful in cases where the researcher is establishing her own baseline model as opposed to comparing her model to one specified by previous researchers.

We propose three criteria for creating predictive benchmarks that can be applied to any social science outcome observed longitudinally. In fact, it is important to point out that a necessary condition for a strong benchmark model should be that it is transportable to different outcomes and not tailored to one specific application. Such transportability can afford the benchmark model and its interpretation in terms of how well our explanatory models explain a given outcome a high degree of consistency across applications.

First, we propose that \emph{the only data to be used in benchmark predictive models should be from the outcome variable.} Not only does this increase the portability of the benchmark model's structure to other outcomes, but, more importantly, an outcome-only model represents the most substantively simple model that can be created. In terms of assembling an explanatory model, an outcome-only model represents the most parsimonious option -- the outcome variable following a self-determining dynamic.\footnote{We want to note an important consideration in building an explanatory model while using a baseline model that includes functions of the lagged dependent variable. The time lag should be specified to either predate or be contemporaneous to any variables in which the researcher is interested in interpreting causally. The identification of a causal relationship can be compromised by conditioning on a post-treatment control variable (i.e., a variable determined after the determination of the causal variable of interest) \citep{Keele2015}. } To create such a model, say of international conflict, we need only to have substantive knowledge of international conflict. Adding covariates complicates the substance of the problem greatly; if we regress conflict on joint democracy, trade, and common IGO membership, we must have substantive knowledge of each of those covariates, confidence in the measures, and understand the processes by which they relate to conflict.

All of this is not to say that an outcome-only model is structurally simple, such models are often quite complicated in their underlying mathematical forms. Aside from the rather obvious effect of previous observations on current observations (e.g. autocorrelation), political science 
is marked by powerful endogenous effects that manifest through networks of interactions and associations, latent or observed, within the outcome variable. 
In other words, without including the complexity of exogenous covariates, there is often much structure that can be included in an outcome-only model. 
For example, recent studies have shown that, in networks, certain endogenous structures are transportable across outcomes \citep{Hanneke:2010, Cranmer:2011, Cranmer:2012a, Cranmer:2012b, Desmarais:2010, Desmarais:2011ieee, Desmarais:2012}.

As a second criterion, we argue that \emph{all predictions must be made strictly out-of-sample.} The reasons for this are simple.  First, in-sample prediction is not \emph{true} prediction because it is predicting observations that have already occurred within the training set. From a statistical perspective, it matters little whether the observations being predicted out-of-sample have yet occurred in nature, but more whether they have yet occurred in the training set. 
Second, in-sample-prediction runs the risk of leaving a model that overfits the data undetected. Overfitting occurs when the statistical model captures artifacts of the dataset (i.e. random error) that are not part of the data generating process. An overfitted model will typically produce good in-sample predictions, but poor out-of-sample predictions because the artifacts of the training set it exploited do not carry over to the test set. As such, developing a model that predicts well in-sample may reflect less of a thorough understanding of the data generating process than a model that predicts well out-of-sample. 

Designs for assessing out-of-sample model fit can take many forms, and may depend upon the format of the data. For a dataset of completely exchangeable observations (e.g., survey data), randomly splitting observations into training and test sets or partitioning the data into $k$ validation sets that are used sequentially both for estimation and testing (i.e., $k$-fold cross-validation) is a common approach \citep{Jensen:2000}.  Cross-validation is a robust and adaptable methodology that has been shown to perform optimally in terms of model selection for several classes of model selection problems \citep[see, e.g., ][]{hall1983,nowak1997,droge1999,van2004}. When the data are organized according to a common dependence structure such as longitudinal/panel data or clusters, whole groupings (e.g., time periods) can be omitted as the hold-out sample \citep{rakotomalala2006}. 

Evaluating a model on an independent test set is not a silver bullet for model selection in a single finite study. It is still possible overfit the data using hold-out methods such as cross-validation. As such, since the test set(s) in any one study have already been exploited to test multiple models, it is important that future studies of the same process either grow the test set(s) or use completely new and independent test set(s). Generalization error is  the prediction error of a model on data from the same population, but outside of the sample.  \cite{cawley2010} show that, even though cross-validation is a nearly unbiased method of estimating generalization error and single-split-sample (i.e. training/test) estimation is unbiased, the variability with which hold out methods estimate generalization error can lead to overfitting in a finite sample. In other words, any one test set may lead to the inclusion of more variables or model components than exist in the true model. \citeapos{cawley2010} results emphasize the importance of accumulatively growing the sample of test data and replicating past studies in order to realize the long run benefits of testing on held out data.

Our last criterion for a good benchmark predictive model is that \emph{the criterion for judging predictive accuracy must be appropriate for the distributional features of the predicted variable.}  This is an important point, but one for which it is difficult to provide general advice. For example, the most relevant feature when it comes to a dichotomous variable is the rarity of ``positive'' events. 
Standard metrics for judging predictive accuracy, the receiver operating characteristic (ROC) curve in particular, can produce misleading results when applied to rare events because this criteria weights the prediction of events and non-events equally. Consider the case of predicting war: war is rare and a (useless) model predicting never war will be overwhelmingly correct. In our application below, we consider this problem specifically and introduce alternative criteria for judging predictive accuracy

\section{Illustrative Application:\\ Predicting International Conflict} \label{sec:application}

We now endeavor to demonstrate as many of the advantages of predictive models discussed above as possible within the confines of a single example. We attempt the prediction of violent international conflict, something notoriously hard to predict, and consider what predictive modeling can teach us about this processes.  

\subsection{The Paucity of Predictive Models of Inter-State Conflict}

Empirical analysis in conflict processes research relies almost exclusively on explanatory modeling, typically using regression. Predictive models, which do not necessarily aim to operationalize a causal theory, are then often seen as the tools of applied scientists or policy analysts rather than of the basic, explanatory science in which we typically engage  \citep{Schneider:2010, Schneider:2011}. It is perhaps not surprising then that there is little predictive work in this field and what does exist is relatively recent. 

\cite{Beck:2000} touched off the contemporary debate on predictive models for conflict with a study that uses a neural network approach, which predicts 17\% of conflicts, compared to 0\% by a conventional logistic regression. This study led to much debate over the utility of restricting samples to only dyads that had a reasonable chance of conflict in the first place, and even sparked some interest in neural networks (which we discuss further below), but failed to produce a substantial literature on predictive models for conflict. In one of the few studies of conflict prediction that followed \cite{Beck:2000}, \cite{Ward:2007} use a Bayesian, Hierarchical, Bilinear, Mixed-Effects model stratified by time to gain an improvement in out-of-sample prediction, again over a fairly standard logistic regression; in this case, the one originally proposed by \cite{Oneal:1999}. The model offers a substantial improvement in predictive ability over logit, but does not compare the performance of its method directly to that used by \cite{Beck:2000}. 

One reason the predictive literature on international conflict is so sparse may be that the structure of the conflict data is such that predictive modeling is difficult with existing technology. For example, time-series approaches to prediction, well established in both economics and political science, are difficult to apply to data that span every possible conflictual relationship in the world over time. None-the-less, there has been a recent increase in predictive work on other conflict processes, including civil wars \citep{Rost:2009, Ward:2010}, transnational terrorism \citep{Desmarais:2011ieee}, and single-conflict time series analyses \citep{Pevenhouse:1999, Schrodt:2000, Brandt:2011, Schneider:2012}.



\subsection{Methods and Measures}

\subsubsection{Predictive Design}
The process of building the predictive models follows that proposed by \cite{Desmarais:2011ieee}. First,  a predictive network is constructed by aggregating conflict initiations over an interval preceding $t$ -- we consider one, five and ten year intervals. Second, for each directed dyad $ij$, a vector of directed dyadic statistics, denoted $\bm{\delta}_{ij}^{t-1}$ are computed on the predictive network. These statistics could include and indicator of whether $i$ initiated a conflict with $j$ during the predictive time interval, a measure of the total conflict activity of $i$ and $j$ during the predictive time interval, or the geodesic distance between $i$ and $j$ in the predictive network. Third, a forecasting model is used to forecast the edge from $i$ to $j$ at time $t$ ($N_{ij}^t$). A simple example could be to estimate the probability that $N_{ij}^t=1$ by $1/(1+\exp(-\bm{\beta}'\bm{\delta}_{ij}^t))$ (i.e., logistic regression). Indeed, \cite{Desmarais:2011ieee} formulate their algorithm using a temporal exponential random graph model (TERGM). However, since the predictive network features are all observed prior to the forecasted edges, their example is a special case of logistic regression \citep{Hanneke:2010}.

\subsubsection{Competing Predictive Algorithms}
As we set about illustrating the abstract discussion above, and evaluating our proposed benchmark model against the performance of the contemporary literature, we must be mindful to ensure that our comparisons are fair. Specifically, we seek to avoid the ``straw man'' comparison in which we apply state-of-the-art predictive methodology and compare our results to those from a well cited paper in the existing literature that did not have prediction as its aim; such a comparison would be unsatisfying if not misleading.  Instead, we illustrate our point by considering three classes of models: those based only on structure endogenous to the outcome measure (the benchmark criteria we proposed above), those based only on exogenous covariates (capturing the large majority of explanatory models observed in the literature), and those that combine both endogenous and exogenous effects. 

We optimize each class of model for predictive performance using one of four classification algorithms: logistic regression, elastic net regularization (i.e., lasso and ridge regression combined) \citep{zou2005}, boosting, and neural networks. These three algorithms are all widely used for classification tasks, and vary both in terms of how the variables are used for classification. We briefly describe each algorithm below.\footnote{The \R packages used to implement elastic net, boosting and neural networks are \texttt{penalized} \citep{penalized}, \texttt{caTools} \citep{caTools} and \texttt{nnet} \citep{nnet}, respectively.} For all of the tuning parameters described below, we set the range of parameter values we test such that none of the optimal parameter values lie at the boundaries of the range.

Elastic net regularized regression \citep{zou2005} is performed by adding two terms to the regression coefficient criterion function (e.g., sum of squared errors, likelihood) that constitute penalties in both the absolute magnitude (i.e. lasso) and squared value (i.e., ridge) of the regression coefficients. The two tuning parameters in elastic net regression are the two weights associated with these penalties. The absolute lasso penalty serves to push a subset of variables that do not contribute enough to the fit of the model to exactly zero. The ridge penalty shrinks regression coefficients towards zero, but does not push them to exactly zero. The elastic net method combines the functions of selection and shrinkage exhibited by the lasso and ridge regression methods of regularization, respectively. We seek to render the algorithms we use comparable in terms of the number of tuning parameters estimated, and therefore set the ridge penalty equal to twice the lasso penalty.\footnote{We chose to fix the ratio of the ridge and lasso parameters due to issues with sensitivity of the lasso penalty. If the ratio exceeded 2 by much, we found that the lasso became ineffective---never selecting down the set of variables. If the ratio were much under 2, the lasso would result in a lack of convergence in the estimation.}

We use feed-forward neural networks with a single hidden layer in the predictive experiments. Neural networks are models that learn some number of functions of the input (i.e. covariate) variables, which then feed forward to predict the outcome. By combining several possibly nonlinear functions of the data, neural networks can approximate the true underlying relationship between the covariates and dependent variable \citep{cybenko1989}.   There are two tuning parameters we consider in the neural network application: the number of nodes in the hidden layer of the neural network, and the regularization (i.e., decay) parameter used to penalize the magnitude of the coefficients linking the variables to the nodes in the neural network.

The boosting methodology we use involves learning one node decision trees based on the covariates. Boosting involves re-weighting the data in iteratively learning weak classifiers. At iteration $t$, a simple classifier is learned with greater weight placed on the data points that are poorly fit in iterations prior to $t$. Combining the iteratively learned classifiers (i.e., decision trees), results in an effective classifier. The important tuning parameter is the number of base/weak classifiers to be learned and aggregated. We follow the \texttt{LogitBoost} methodology proposed by \cite{Dettling:2003}.


\subsubsection{Outcome Variable}
The outcome variable, on which we measure the performance of our predictive algorithms, is the directed network of conflict initiations, $N$, aggregated over a one year interval. The edge from state $i$ to $j$ at time $t$ is one if $i$ initiated a conflict with $j$ during year $t$ and zero otherwise. We only generate conflict forecasts for directed dyads composed of states that were both in the state system in the previous year because our model is not optimized to forecast the entry of new states into the system (a process usually unrelated to the occurrence of \emph{international} conflict). We use the Correlates of War dataset (v3), which covers 1816--2001, and we focus on outcome years 1979--2001.  

\subsubsection{Performance Criteria}
We propose that the ROC curve is a fine criterion when analyzing data that display a good balance between events and non-events (such as voter turnout), but that the area under the precision recall (AUC-PR) curve is better suited to the analysis of rare events. Consider the four types of predictions one can make for a binary variable, displayed in the left cell of Table \ref{criteria} based on \citet{Davis:2006}: true positives (TP), false positives (FP), false negatives (FN), and true negatives (TN). Many metrics can be computed from the quantities in the contingency table, but three are of particular interest here: the precision, the true positive rate or recall, and the false positive rate. These metrics and their computation are displayed in the right cell of Table \ref{criteria}. The ROC curve, familiar to most subfields of political science, plots the false positive rate on the $x$-axis and the true positive rate on the $y$-axis. Conversely, the PR curve plots recall (the true positive rate) on the $x$-axis and the precision on the $y$-axis, thus focusing on the predictions of those positive events that did occur. Similar considerations are necessary when dealing with count, continuous, or other types of variables.\footnote{Replication data are posted to the {\em Political Analysis} Dataverse \citep{replication_data}.} 

\begin{table}
\flushleft
\begin{tabular}{cc}
\parbox{72mm}{
\begin{tabular}{r|c|c|}
& actual & actual \\
& positive & negative\\ \hline
predicted positive & $TP$ & $FP$ \\ \hline
predicted negative & $FN$ & $TN$ \\ \hline
\end{tabular}
}
&
\parbox{10mm}{
\begin{tabular}{rcl}
Precision & = & $\frac{TP}{TP + FP}$ \\
True Positive Rate (a.k.a. Recall)& = & $\frac{TP}{TP + FN}$\\
False Positive Rate & = & $\frac{FP}{FP + TN}$\\
\end{tabular}
}
\end{tabular}
\caption{Common metrics for judging predictive accuracy. The left cell shows the contingency table and the right cell shows the metrics of interest.}
\label{criteria}
\end{table}

The challenge, when deciding between the ROC and PR curves, can involve either the rarity of the subject under study or the importance of accurately predicting events vs non-events. We argue here, and will show with application below, that the ROC curve is generally inappropriate when examining rare events. The reason being that the ROC curve weights the value of accurately predicting non-events (zeros) the same as accurately predicting events (ones). This is often a desirable attribute of the ROC, but consider the effects of this weighting when analyzing rare events. For example, out of the more than one million bilateral wars that \emph{could have} happened since 1816, less than 1,000 have. As such, a model that never predicts war would do well by the ROC criterion because it is good at predicting the modal category of no war. In terms of the measures presented in Table \ref{criteria}, AUC-ROC values look high simply because the false positive rate is nearly always extremely low due to $TN$ in the denominator including nearly every case in the data.

\begin{figure}
\begin{center}
\begin{tabular}{cc}
AUC-ROC & AUC-PR\\
\includegraphics[height=6cm]{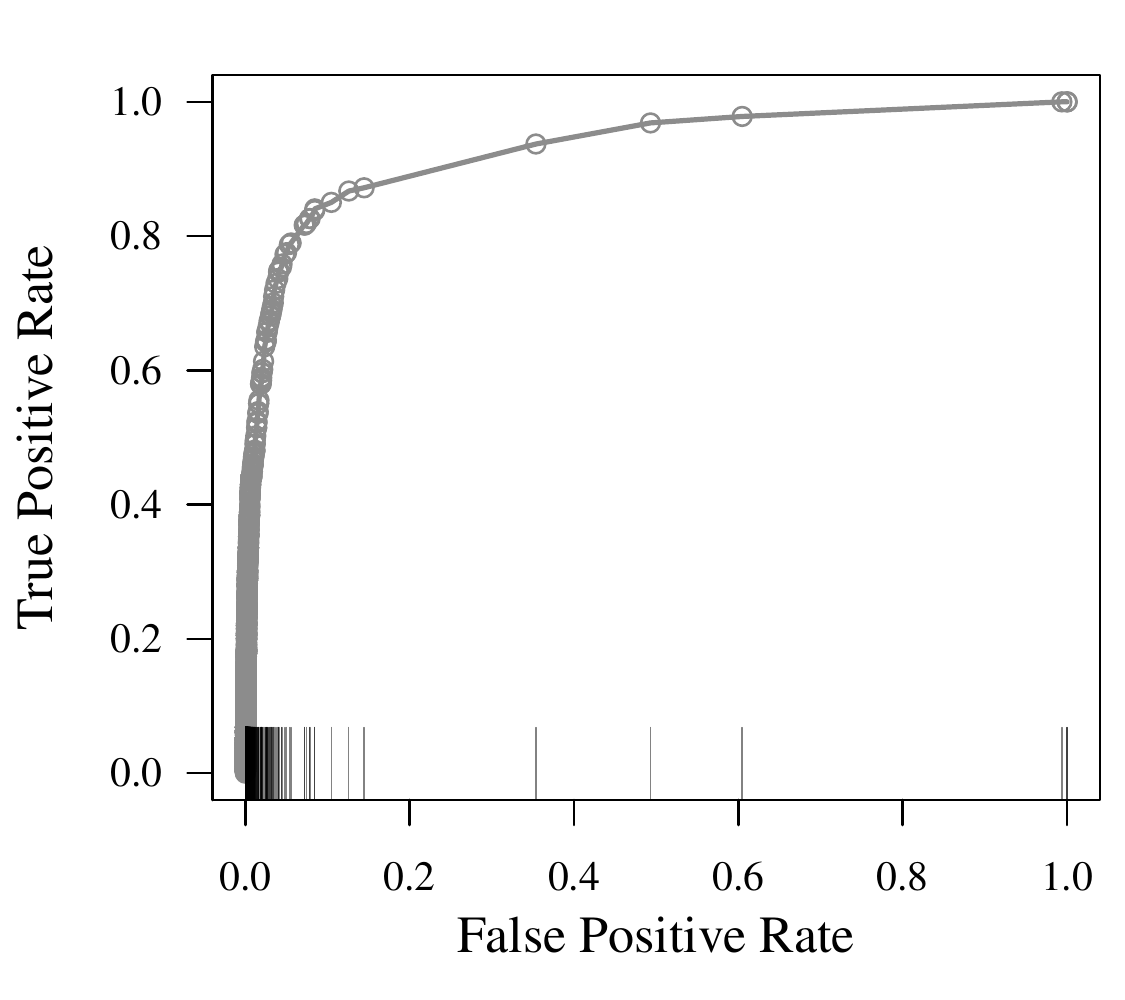} & \includegraphics[height=6cm]{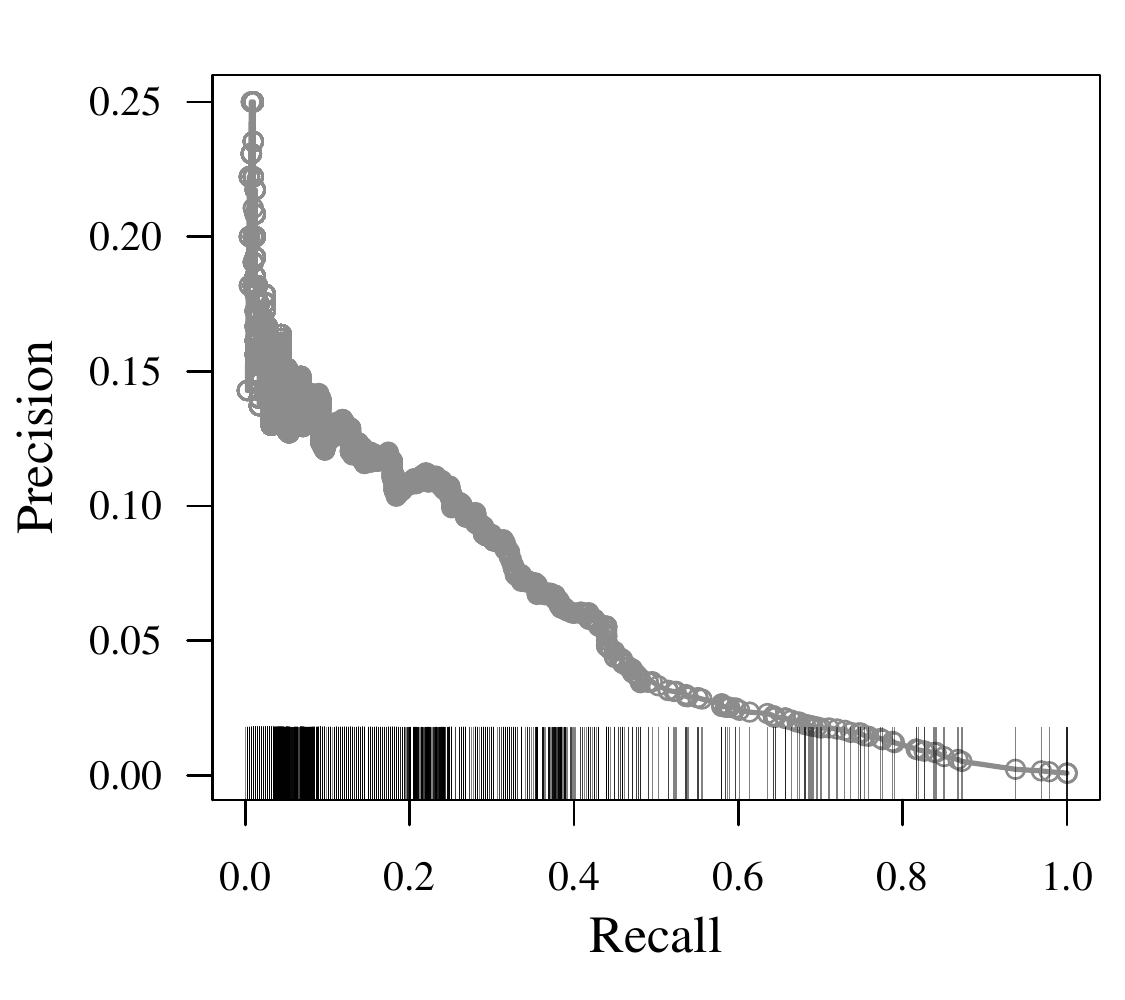} \\
\end{tabular}
\end{center}
\caption{Example data points from which AUCs are constructed. Each point represents a predicted probability of war produced by at least one of the directed dyads in the most effective model we find in the modeling exercise in Section \ref{sec:application}.}
\label{AUCs}
\end{figure}

In contrast, AUC-PR does not involve $TN$. 
Figure \ref{AUCs} provides a graphical example of how AUC-PR better represents the predictive task than AUC-ROC for rare events. It depicts the results from our best performing war prediction model found in the analysis below. In both the PR and ROC curves, points move from left to right on the plot by lowering the predicted probability threshold at which a positive is predicted. From the AUC curve, it can be seen that the FPR is small for all but the smallest thresholds, and the vast majority of the area under AUC resides under a curve extrapolated over fewer than ten cases. Because the PR curve does not involve $TN$, there is much less skew in where the points reside on the curve.  Additionally, because the height of the curve decreases with the $x$-axis, the majority of the area under the PR curve resides under the bulk of the points observed in the data, lending greater confidence in the certainty about the value of AUC-PR.\footnote{We use the function \texttt{auc.pr} in the \R package \texttt{minet} \citep{minet} to calculate AUC-PR and the \texttt{performance} function in the \texttt{ROCR} package \citep{ROCR} to calculate AUC-ROC.} Furthermore, in a recent simulation-based comparison of AUC-ROC and AUC-PR, \cite{ozenne2015} show that AUC-PR performs more effectively than AUC-ROC in selecting diagnostic biomarkers in rare diseases.

\subsubsection{Measures}

There are three classes of predictive measures in our analysis -- those that rely solely on past iterations of the  conflict network to predict conflict at time $t$ (endogenous network measures), and those that are built using additional data sources (exogenous covariates) and three that are something of a hybrid between endogenous and exogenous, we call these semi-endogenous. We present the components of these classes.

In keeping with our proposed benchmark criteria, in which the benchmark model is created using only the outcome variable, we consider several measures endogenous to the conflict network. 
All of these endogenous measures are designed capture the similarity of any two nodes in a network in terms of certain specific network structures. Networks offer a representation that permits the extension of endogenous effects beyond the individual dyad. We draw upon the networks framework to formulate the best possible candidate model(s) of endogenous effects. As such, the endogenous effects we include can be thought of as proximity measures, in network space, between any two states in the system of interlocking conflicts. We follow \cite{Desmarais:2011ieee} in our selection of endogenous effects. 
These effects include the following.

\begin{itemize}
\item {\bf Flow}: The product of the number of conflict initiations sent by the prospective conflict sender in a dyad, and the number of conflict initiations received by the prospective recipient of conflict in a dyad. 
\item {\bf Common Community:} An indicator of whether two states were in the same community, as determined by an algorithm for community detection in networks. Community detection algorithms partition the network into sets of actors that tie with each other at a much higher rate than would be expected based on the number of ties to and from each actor. We use the ``walktrap'' community detection algorithm \citep{Pons:2005}.
\item {\bf Common Combatants}: The number of third states with which both states in the dyad went to war within the lagged interval.
\item {\bf Adamic-Adar Similarity}: Similar to common combatants, but each third state is added to the count of common combatants with a weight that decreases logarithmically with the number of other connections held by that third state (i.e., adjusting for the intuition that sharing a partner that itself has many other ties may not indicate proximity between two states) \citep{adamic2003}. 
\item {\bf Jaccard Similarity}: Number of common combatants divided by the total number of unique states to which at least one of the states in the dyad is connected. This measure accounts for the tendencies for the two states in the dyad to form ties\citep{leicht2006}. 
\item {\bf MMSBM}: Probability of a tie between two states using a mixed-membership stochastic blockmodel fit to the network in the lagged interval. In a block model, each actor is attributed with a latent class (or block). The blocks are defined by probabilities of interaction with all other blocks. A block model is a simple latent class model for predicting ties in networks. The mixed membership variant allows actors to be in each class with varying probabilities \citep{airoldi2009}. 
\item {\bf Latent Space Distance:} The latent space model for networks is another latent variable model for fitting the probability of a tie between two actors based on network structure. Similar to ideal point analysis, each actor is attributed with a latent position in two-dimensional space. The probability of a tie between two actors is then inversely related to the Euclidean distance between actors in this space \cite{Hoff:2002}. We fit latent space models to the lagged networks and include distances between states in the lagged latent space to predict conflict.
\end{itemize}

We are not the first to propose that inter-temporal dependencies should play a major role in models of interstate conflict. \cite{beck1998} show that the likelihood of conflict between two states at time $t$ depends upon the status of the dyad going back up to 10--15 years, with recent conflict predicting a relatively high likelihood of current conflict, and the likelihood of conflict decaying with the number of years of peace in the dyad. \cite{dafoe2011}, in a replication of \cite{gartzke2007}, demonstrates that accurately modeling temporal dependence is important to identifying relationships between conflict and state attributes, such as the ``democratic peace''. The endogenous effects we propose specifying above can be seen as an extension of this earlier work on within-dyad temporal dependence. That is, we hypothesize that the status of the conflict relationship between states $i$ and $j$ at time $t$ depends not only on the history of conflict between those two states, but also on features of the historical positions of $i$ and $j$ in the broader conflict network.

Lastly, we include a series of exogenous covariates that are commonly used in the conflict literature. While not an exhaustive list, these covariates operationalize many of the major theories of interstate conflict and, at minimum, represent the set of usual controls for such studies. These covariates are all measured at the commonly used dyadic level and include joint democracy, trade dependence, joint IGO membership, CINC (power) ratio, and the geographic distance between capitols, as well as indicators for whether the dyad includes at least one major power, a defensive alliance, or physically contiguous states.\footnote{Contiguity, IGO and CINC data come from the correlates of war project \citep{singer1972,pevehouse2004,stinnett2002}. Trade data come from \cite{gleditsch2002}. Distance data come from \cite{gleditsch2001}. Joint democracy is derived from Polity IV scores \citep{marshall2002}.}

\subsection{Results}

We seek to illustrate the several virtues of predictive modeling discussed theoretically above. We do so by considering each predictive virtue in turn. But first, we consider the three rival machine learning techniques so that we can focus subsequent discussion on the best performer. 
In Figure \ref{allBar} we report the average areas under the PR and ROC curves for all of the estimation algorithms and model specifications. Averages are taken over years, with each year constituting a single observation. We also computed nonparametric bootstrap 95\% confidence intervals for each average by simple bootstrap resampling of means. Due to the complex dependence among dyads within years \citep{Cranmer2016}, there is no straightforward way to resample dyads to construct bootstrap confidence intervals. As such, the confidence intervals we construct, which are based on resampling from twenty three years, are fairly conservative.  

We see in Figure \ref{allBar}, that the elastic net performs better on the whole than the boost, which in turn performs better on the whole than the neural network. While this result may be surprising given that the elastic net  is the simplest of the algorithms we applied, we also find it intellectually appealing in its simplicity. One limitation of this result is that, especially when it comes to the area under the PR curve, the bootstrap confidence intervals do overlap from one algorithm to another, so it is not clear that the performance levels of the algorithms are statistically distinguishable. One characteristic of our predictive models that can be discerned given the averages and CIs in Figure \ref{allBar} is that the areas under the PR curve for our models significantly exceed that which would be expected based on a random ordering of the dyads. If the predicted probability of a positive is drawn uniformly at random, the expected area under the PR curve is simply the rate of positives in the test data (which is well under 1\% in all of the years in our data) \citep{lopes2014,esteban2015}.

\subsubsection{Quality of the Benchmark Model and the Current State of Knowledge}

\begin{figure}
\centering
\begin{tabular}{c}
\includegraphics[width=15cm]{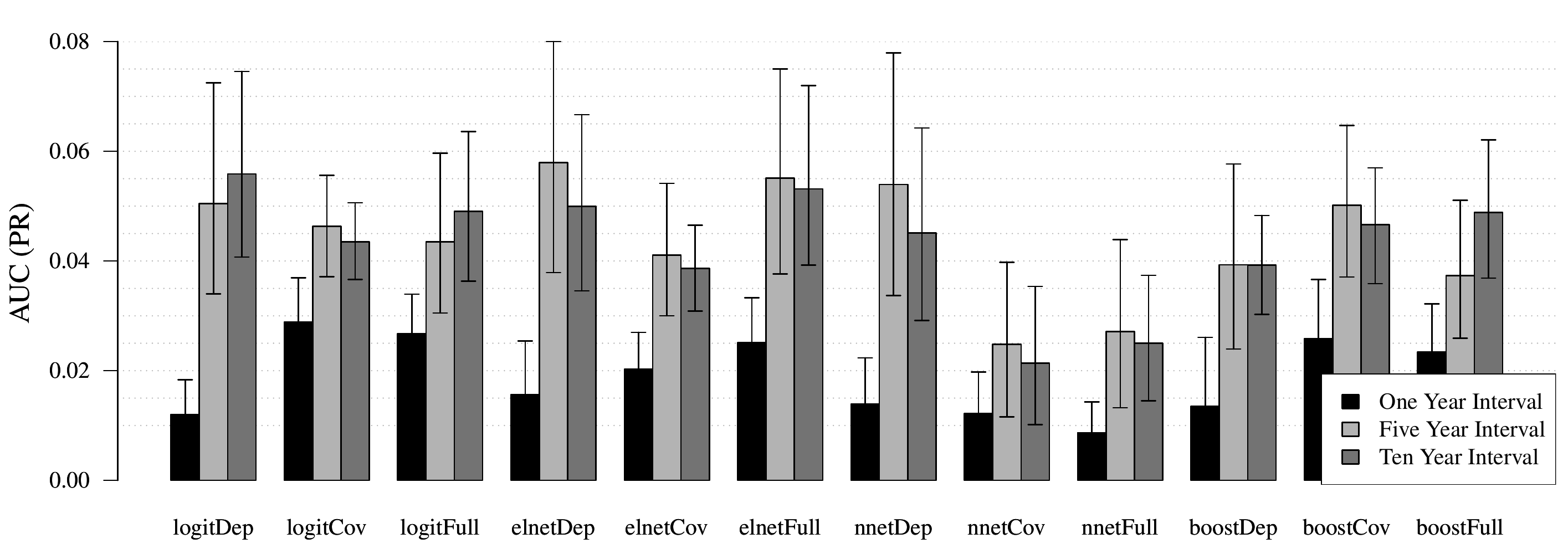}\\
\includegraphics[width=15cm]{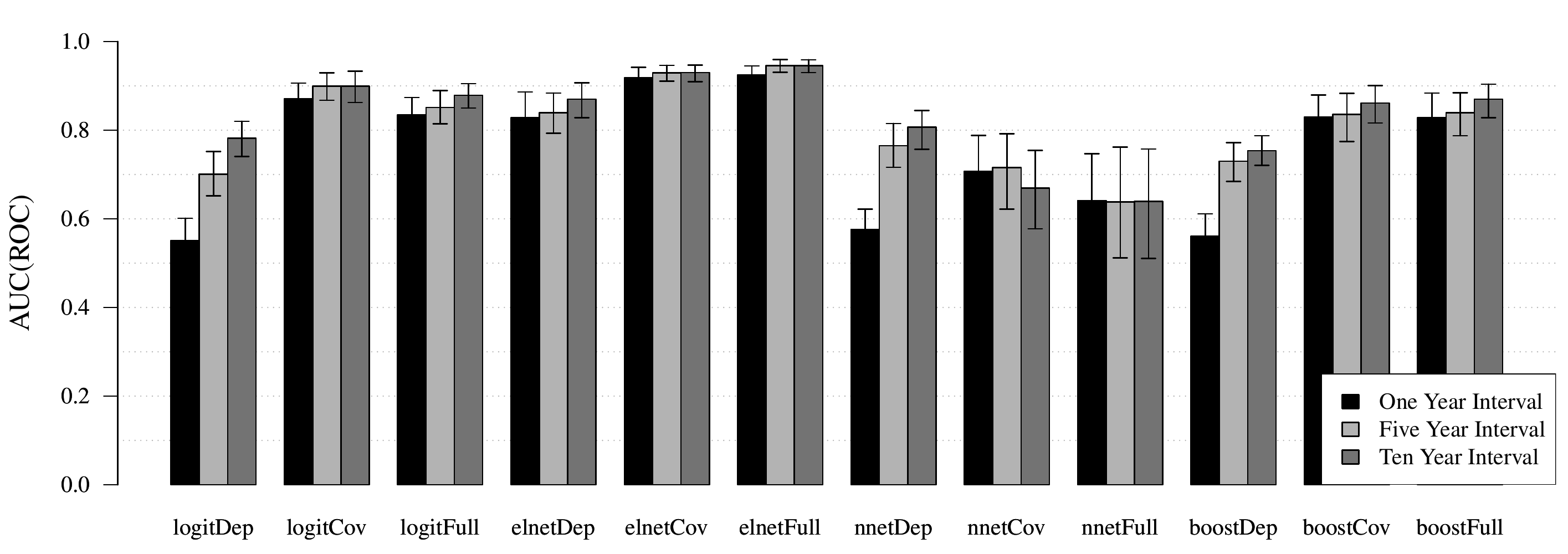}
\end{tabular}
\caption{Comparison of alternative predictive models by different criteria. Both plots show each of the predictive models, respectively using the dependent variable only, covariates only, and both together with one, five, and ten year training intervals. The upper plot uses the area under the PR curve as the fit criteria, the lower plot uses the area under the ROC curve.} 
\label{allBar}
\end{figure}

The most interesting result apparent in Figure \ref{allBar} is that the somewhat naive benchmark model is the single best predictor: the elastic net model with network statistics only and a five year training window is the best model our analysis was able to produce. 
The facts that the best model is outcome-only and that predictive performance is usually decreased when exogenous covariates are added to the outcome-only benchmark model has troubling implications. This result suggests that the vast majority of the literature on international conflict, which has not accounted for endogenous network effects, has missed the dominant predictive attributes of the conflict process entirely. Also, that predictive power decreases when covariates are added to the network structure suggest that including both constitutes overfitting the training data, a troubling implication of the use of covariates indeed.

A second major result reflected in Figure \ref{allBar} is that the maximum predictive accuracy any of our models were able to achieve as slightly over 7\% by the PR criterion. Substantively, this means that our strongest model accurately predicts approximately 7\% of those conflicts that ultimately occur (recall that the PR curve focuses on the accurate prediction of events and not the accurate prediction of non-events). This is rather less than one would have hoped given the long history of the study of international conflict. This suggests one of two things. First, it is possible that the level of noise-to-signal in the conflict data is quite high, making accurate predictions difficult with current measures. In other words, there is so much stochastic variation in the data that 7\% is the best we can do under the benchmarking rules we set for ourselves and with available measures, even though our theories accurately capture the causal process underlying international conflict. That is to say, even though we may know the data generating process, the process is chaotic and we lack the ability to measure inputs and starting values with sufficient precision to predict (as was the case considered above with coin flipping). The alternative interpretation is that the elephant is in the room, but we have not seen it. In other words, there is, missing from our current understanding of conflict processes, a (or the) major determinant of violent international conflicts. Most likely, the truth lies somewhere in between, but we none-the-less see this result as cause for deep reflection on the state of our science and how it may be improved. 


\subsubsection{Identification of New Relationships and Dynamics}

First, let us consider the ``memory'' of the process. 
One thing we notice in Figure \ref{allBar} is that the predictive performance of longer training periods (five and ten years) is nearly universally superior to one year training periods (the sole exception being the neural network covariates only model as judged by the problematic ROC), and usually by a wide margin. It is also notable that, in many cases, the five year training window performs better than a ten year training window, or at least similarly. These two results, taken together, suggest that much is gained by having a memory in excess of one year, but comparatively little is gained by jumping from five to ten. This result, that conflict is a long-memory process, has some troubling implications for applied work on international conflict. Often times, a one-year lag of the outcome variable is included on the right hand side of a regression in order to control for temporal dependencies. This result suggest that this practice is generally inadequate for those purposes, and that lags of at least five years should be considered in order to achieve the desired effect. For a detailed discussion of how to choose lag lengths, see \cite{Cranmer:2015}.

\begin{figure}
\centering
\begin{tabular}{rcc}
\begin{sideways} \textbf{\hspace{40pt} One Year} \end{sideways} & 
\includegraphics[width=7cm]{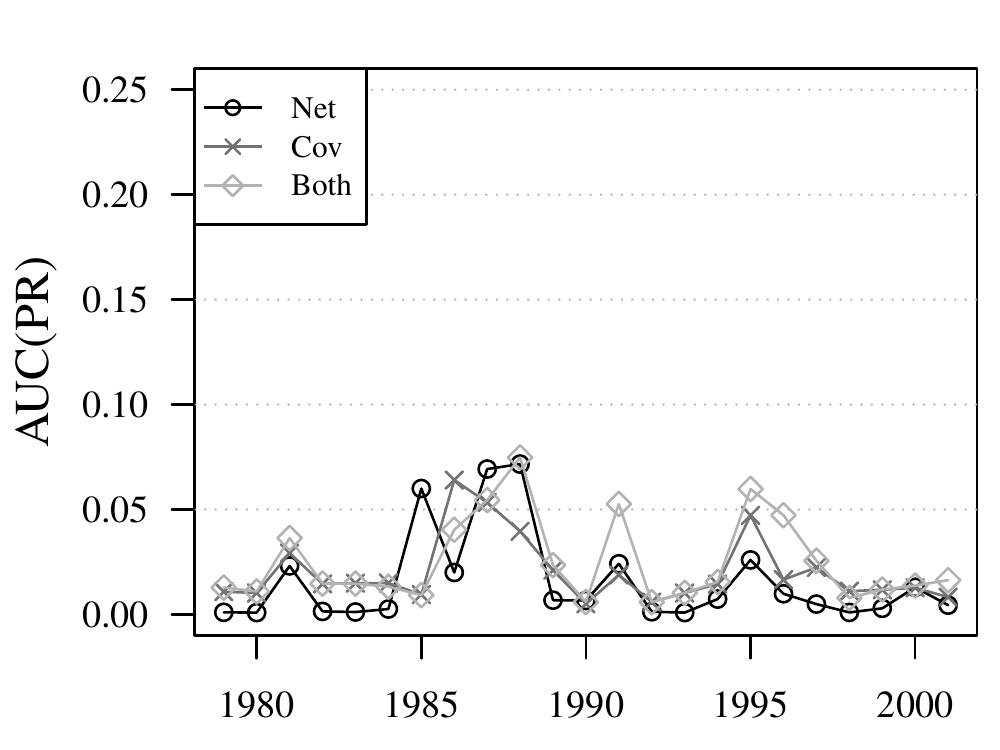} &
\includegraphics[width=7cm]{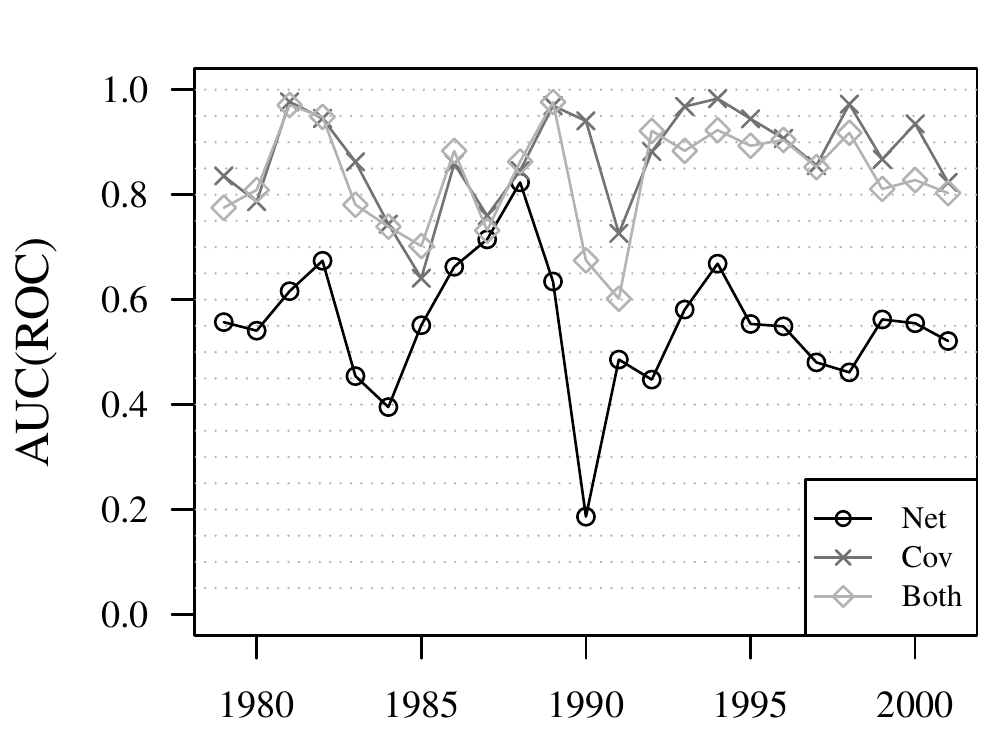}\\
\begin{sideways} \textbf{\hspace{40pt}Five Years} \end{sideways} & 
\includegraphics[width=7cm]{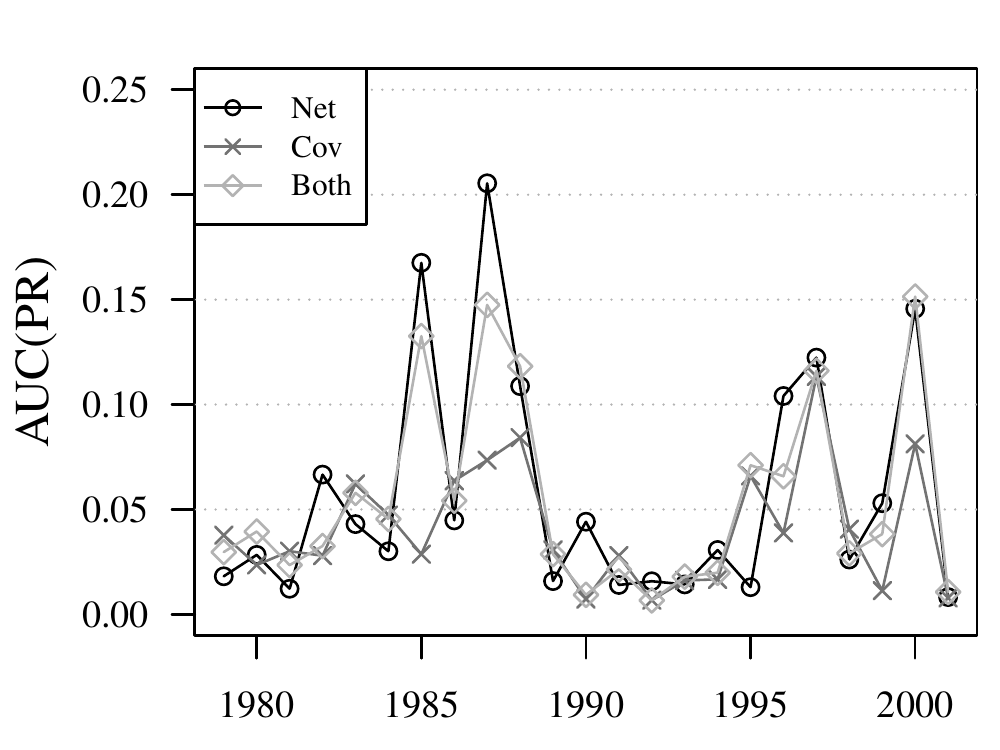} &
\includegraphics[width=7cm]{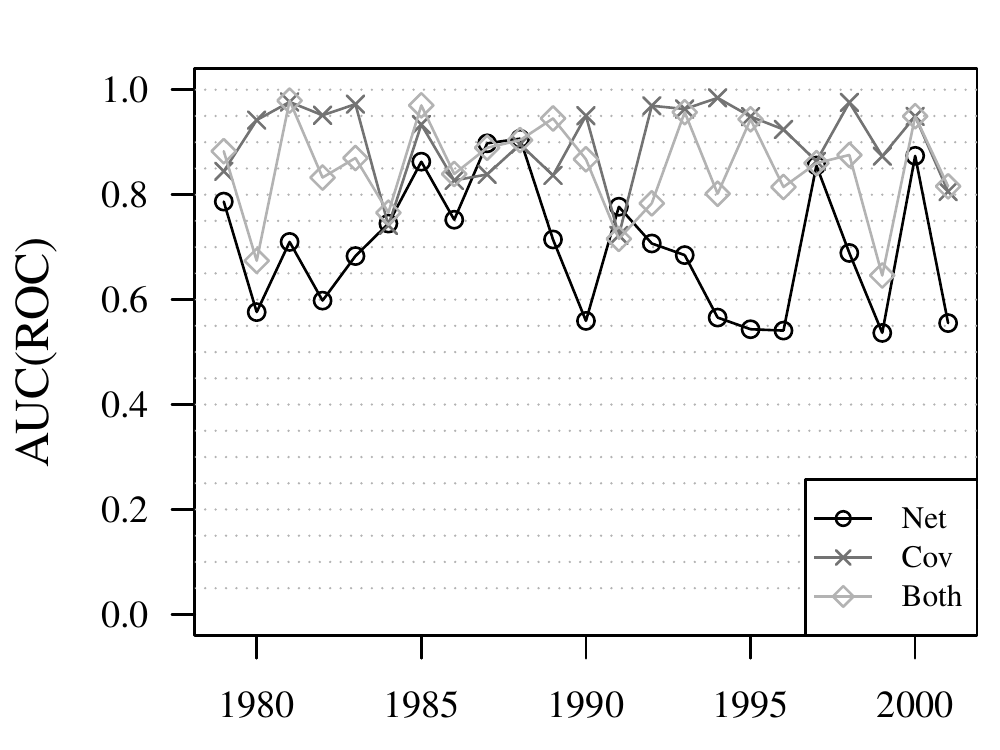}\\
\begin{sideways} \textbf{\hspace{40pt}Ten Years} \end{sideways} & 
\includegraphics[width=7cm]{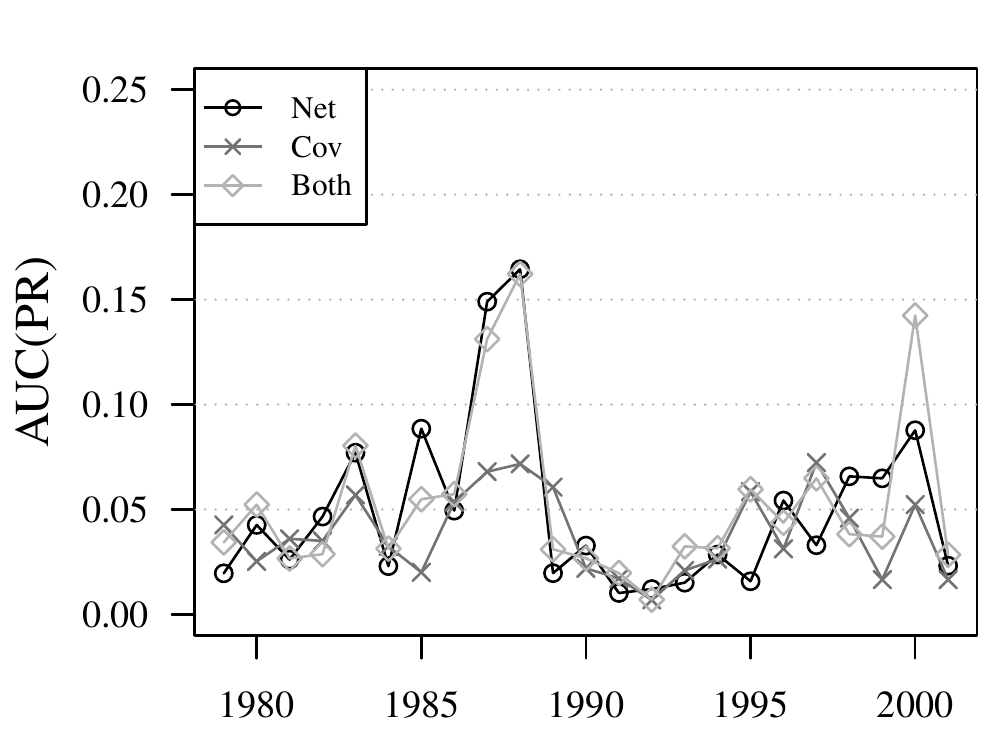} &
\includegraphics[width=7cm]{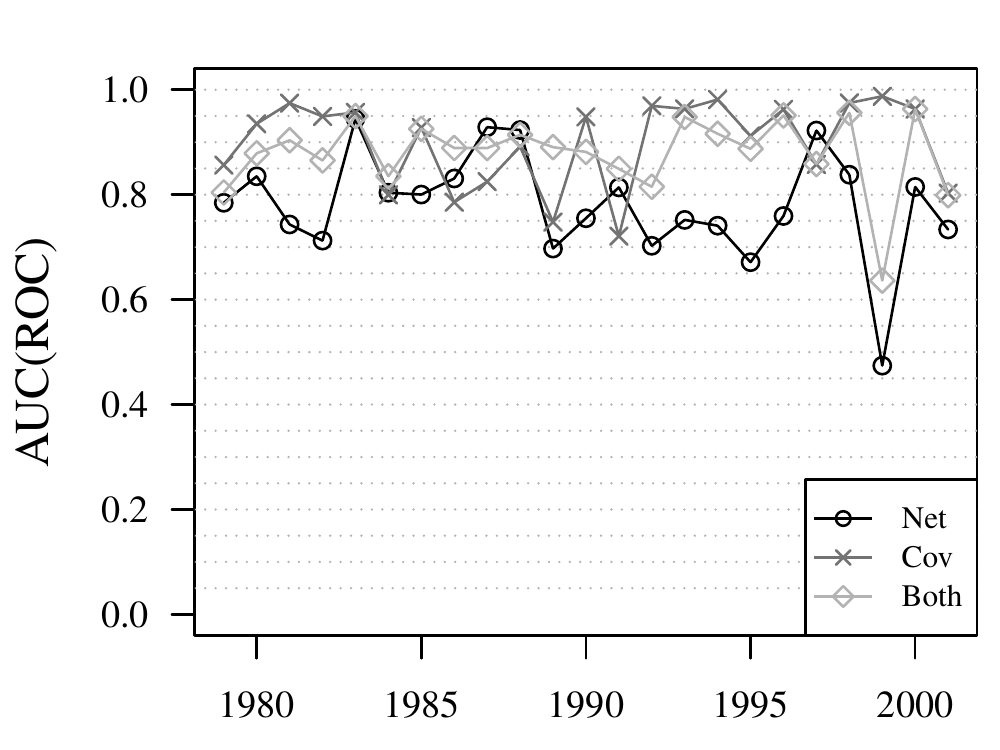}\\
&\textbf{PR} & \textbf{ROC}
\end{tabular}
\caption{Predictive performance for the elastic net over time. Performance is depicted, in terms of both PR and ROC, for each of the three specification types, over 1979--2001.}
\label{lassoOT}
\end{figure}

We can understand the temporal dynamics in greater detail by examining the year-to-year predictive performance of the elastic net models depicted in Figure \ref{lassoOT}, one sees that either the outcome-only model (black line) or the combined outcome-and-covariates model (light gray line) tend to have similar levels of predictive accuracy and both consistently outperform the covariates only model. This provides further, and temporal,  visualization of the result reported above, that the covariates only model, more traditional in empirical international relations, consistently performs the worst out of the three options. Regarding the dynamics, we can see that there is considerable year-to-year volatility in the predictive performance of each model. However, the models that include network dependence exhibit a handful of years in which they perform particularly well at predicting conflict initiations. The covariate-only models do not exhibit comparable up-swings in predictive performance. This distinct dynamic can be considered in contrast to a consistent difference between the covariate only and dependence term models. 

Figure \ref{lassoOT} also shows that the model, whether using a one, five, or ten year training period seems to perform at its best around the mid 1980's and late 1990's/2000. This is interesting because it seems to capture the dynamics of the Cold War, drop off a bit in the immediate aftermath of the Cold War, and then, after some re-training on the differently pattered data, to do well in the contemporary era. 

We further show, for the purpose of illustration, that the more traditional ROC curve, whose shortcomings when applied to rare events we discussed above, provides deceptively promising results for these same analyses, suggesting that we predict something on the order of 90\% of the data. We also see an inversion, when considering the ROC curve, between the predictive performance of the outcome-only network model and the covariates only model; the latter consistently doing better than the prior over the range of data considered.




\subsubsection{Judging the impact of one variable}

\begin{figure}
\centering
\begin{tabular}{cc}
Joint Democracy & War with Ally\\
\includegraphics[width=7cm]{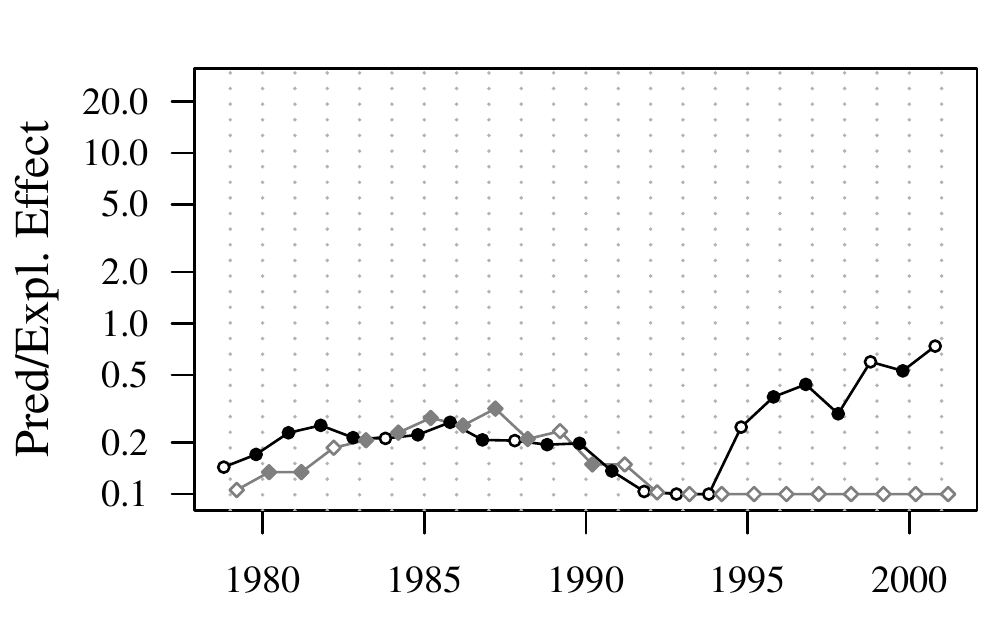} &
\includegraphics[width=7cm]{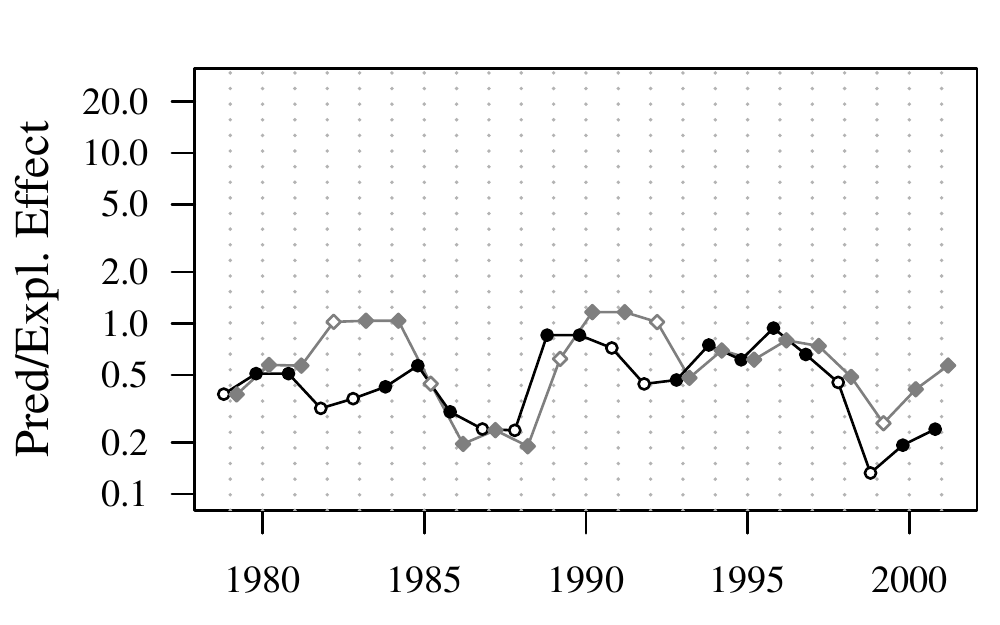} \\
Shared IGO Membership & CINC Ratio\\
\includegraphics[width=7cm]{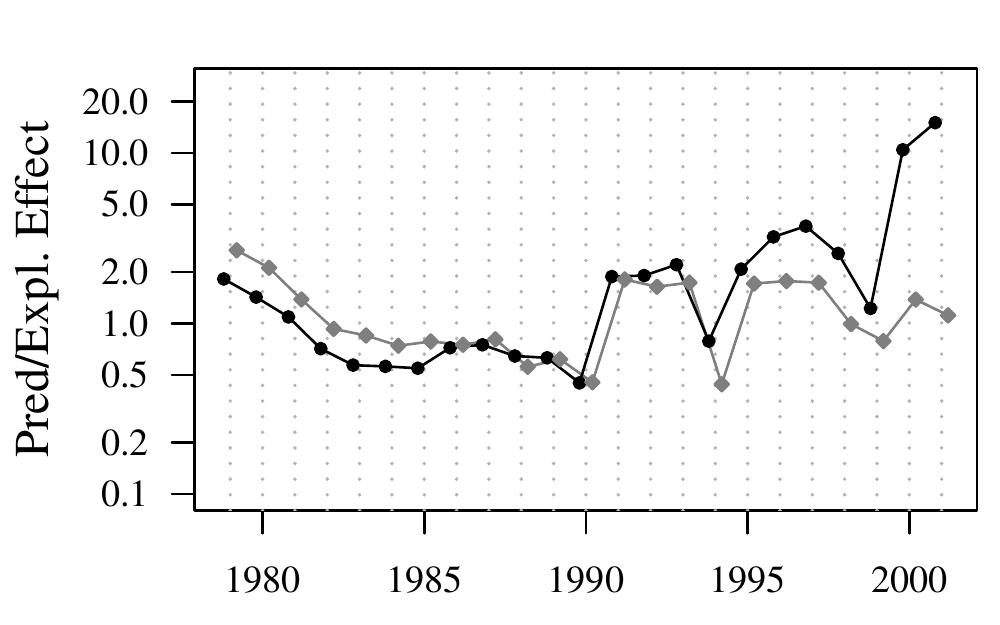} &
\includegraphics[width=7cm]{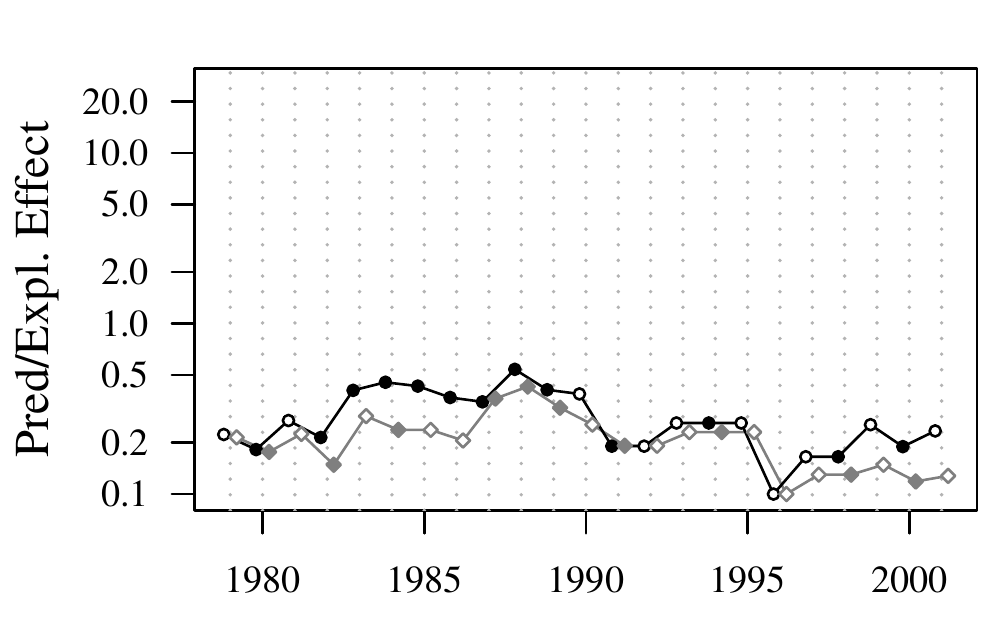} \\
Major Power Dyad & Defensively Allied \\
\includegraphics[width=7cm]{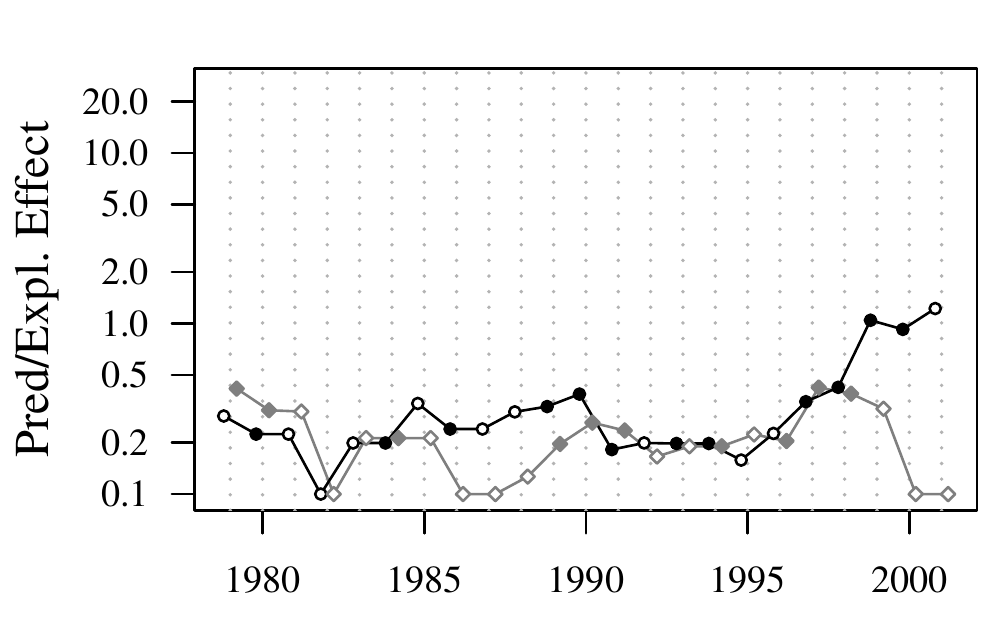} &
\includegraphics[width=7cm]{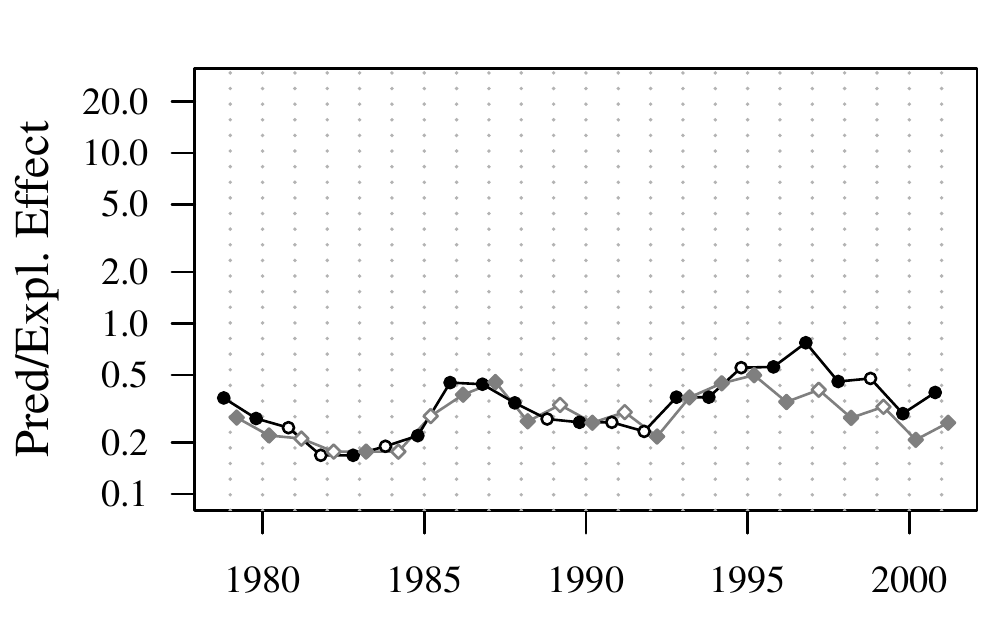} \\
\end{tabular}
\caption{Variable effects as measured by abs(elastic net $\beta$)/abs(logit $\beta$). When this quantity is  high, elastic net has not penalized the coefficient down and the variable can be said to be a stronger contributor to predictive performance. The ratios from the models based on a five-year lagged interval are in black, and those based on a ten-year lagged interval are in gray. To smooth the lines, we depict rolling means over a three-year period centered at the focal year. The points that are not shaded reflect years in which the coefficient ratio fell below 0.01, indicating that the variable was effectively removed from the model through regularization.} 
\label{releffects_covariates}
\end{figure}

\begin{figure}
\centering
\begin{tabular}{cc}
Common Community & Geographic Distance\\
\includegraphics[width=7cm]{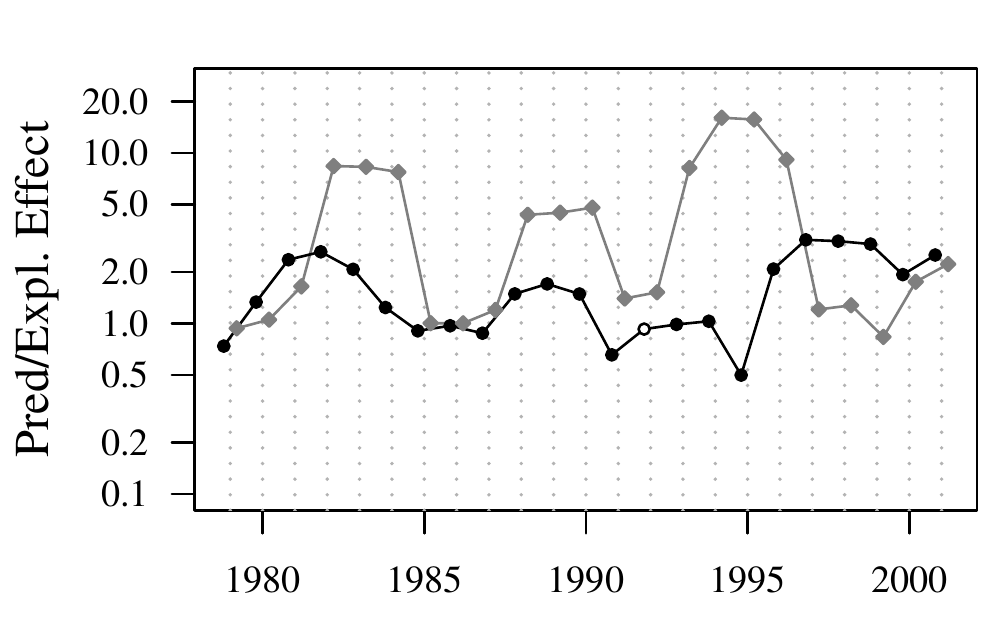} &
\includegraphics[width=7cm]{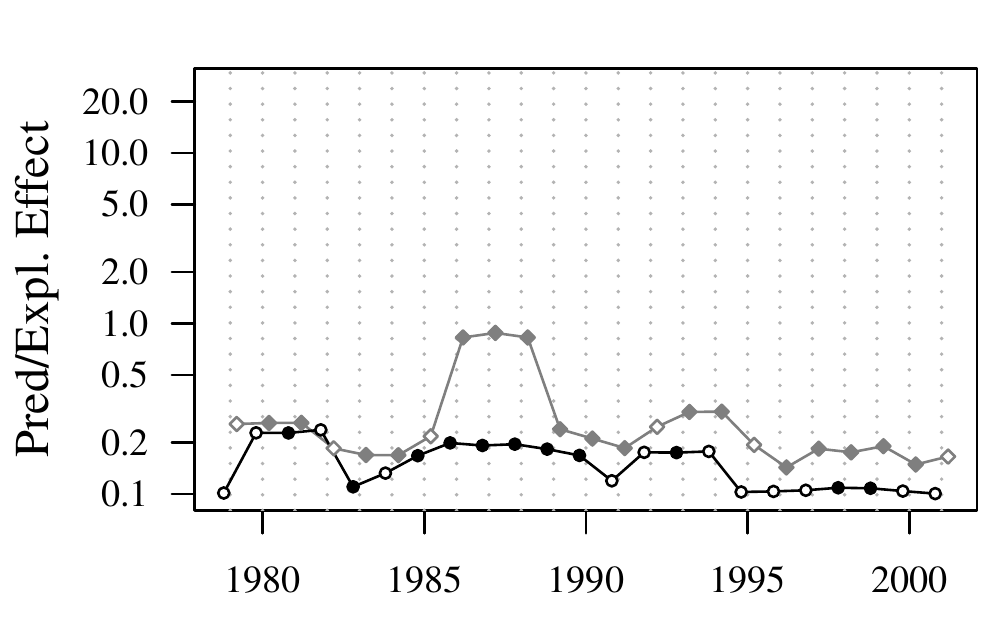} \\
Contiguity &Trade Dependence\\
\includegraphics[width=7cm]{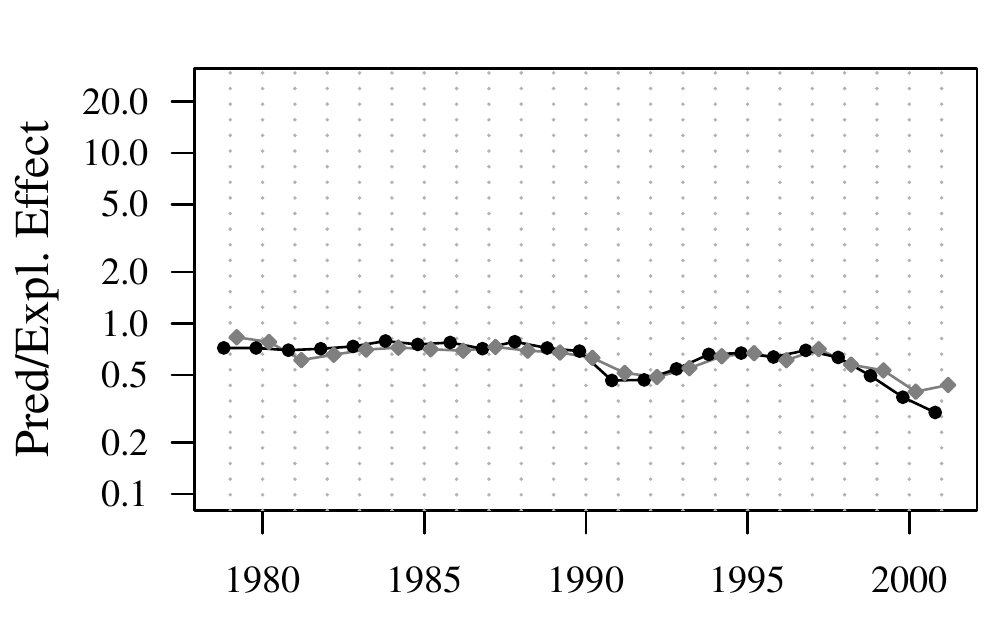} &
\includegraphics[width=7cm]{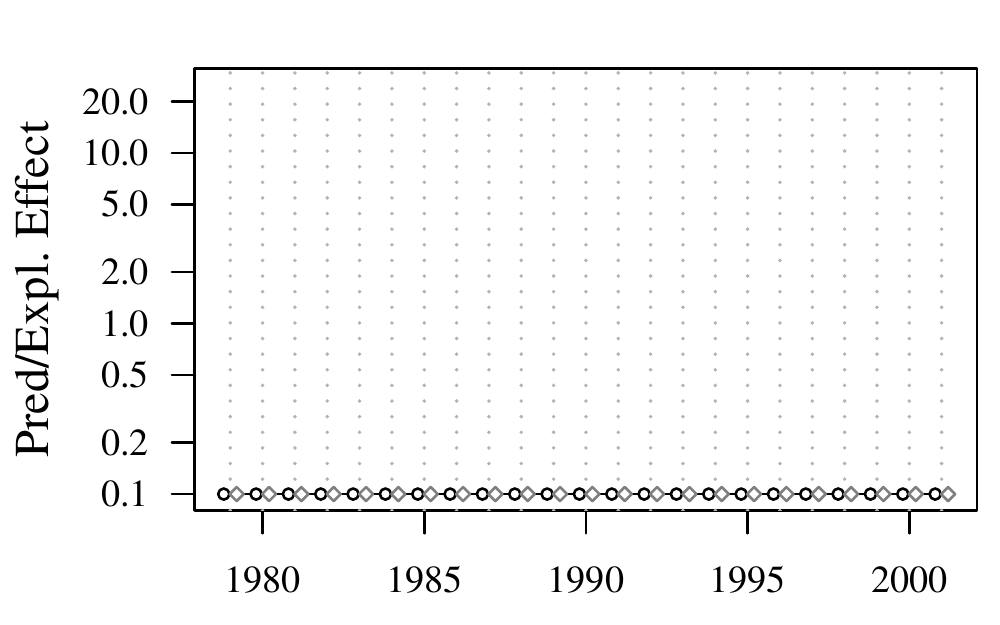} \\
Memory & Flow\\
\includegraphics[width=7cm]{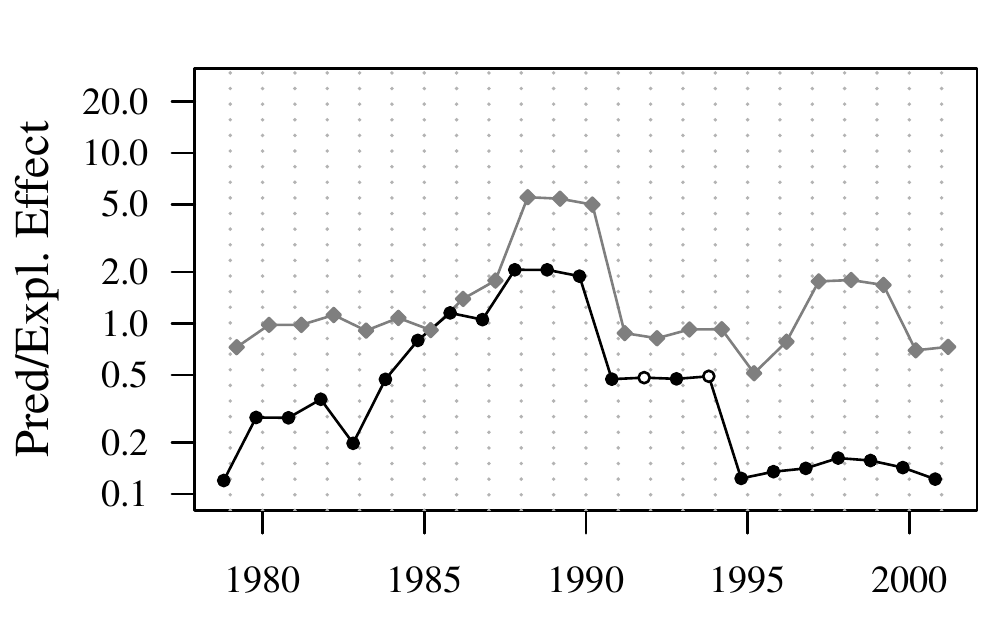} &
\includegraphics[width=7cm]{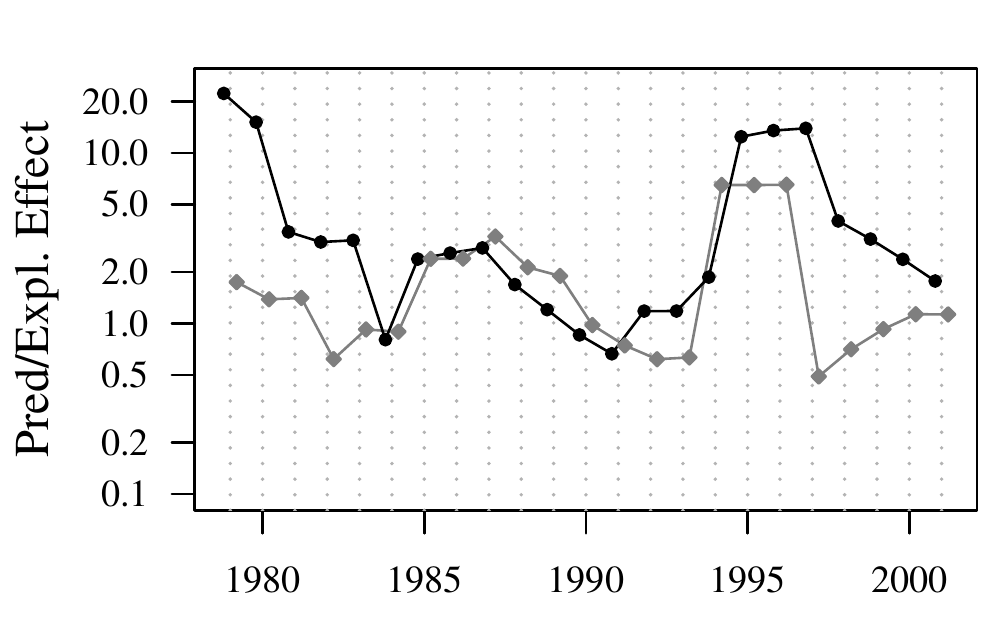} \\
\end{tabular}
\caption{Variable effects as measured by abs(elastic net $\beta$)/abs(logit $\beta$). When this quantity is  high, elastic net has not penalized the coefficient down and the variable can be said to be a stronger contributor to predictive performance. The ratios from the models based on a five-year lagged interval are in black, and those based on a ten-year lagged interval are in gray. To smooth the lines, we depict rolling means over a three-year period centered at the focal year. The points that are not shaded reflect years in which the coefficient ratio fell below 0.01, indicating that the variable was effectively removed from the model through regularization.}  
\label{releffects_mixed}
\end{figure}

\begin{figure}
\centering
\begin{tabular}{cc}
Common Combatants & Adamic-Adar Similarity\\
\includegraphics[width=7cm]{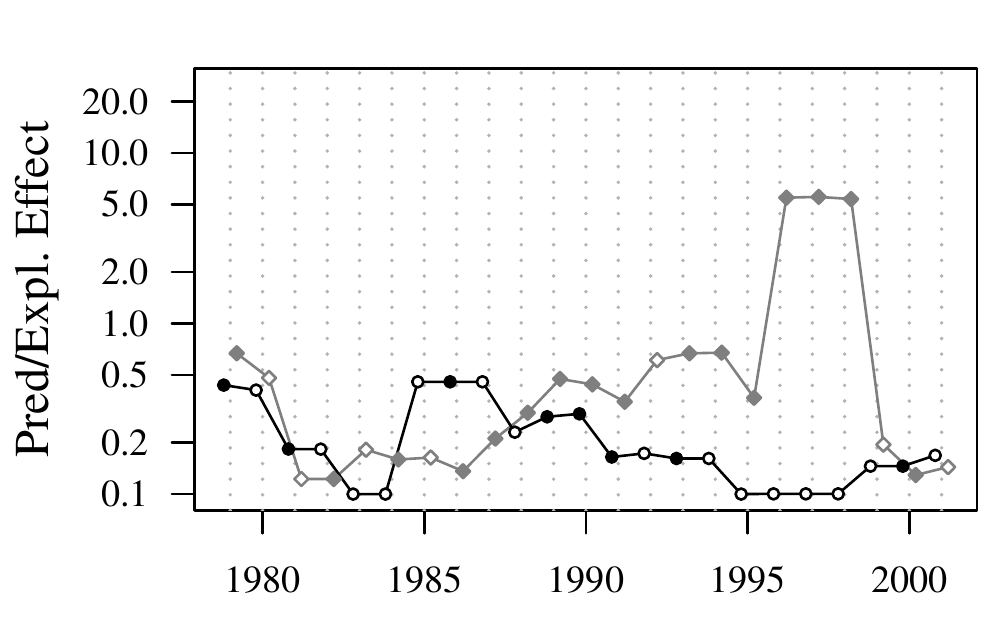} &
\includegraphics[width=7cm]{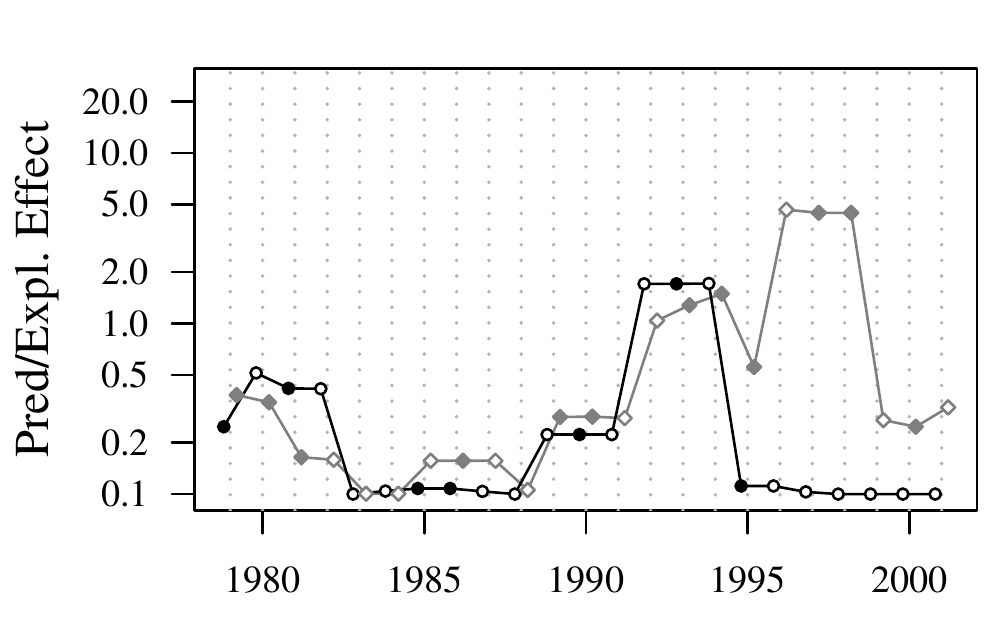} \\
Jaccard Similarity & MMSBM\\
\includegraphics[width=7cm]{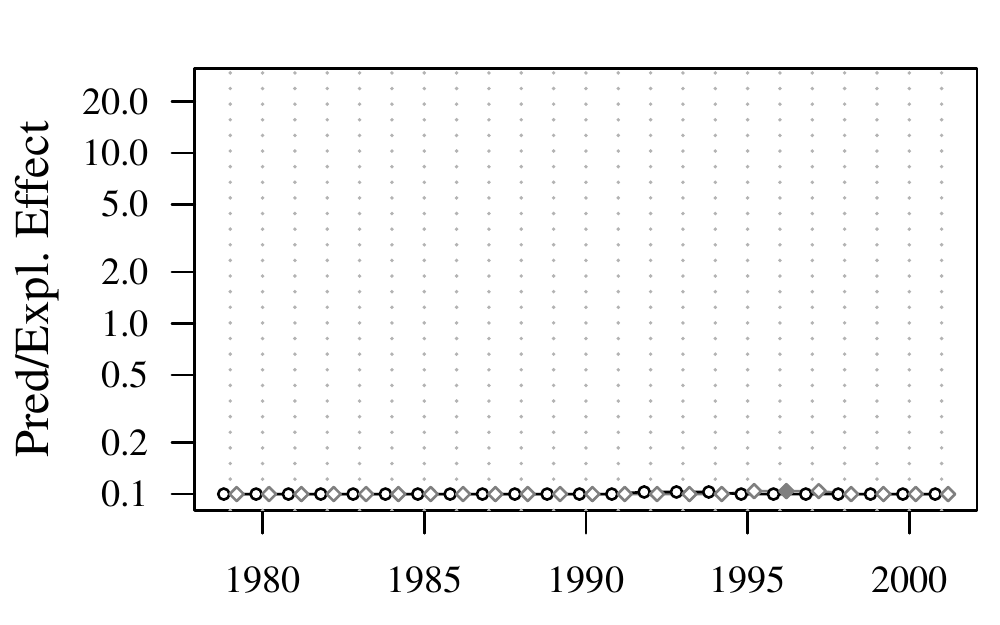} &
\includegraphics[width=7cm]{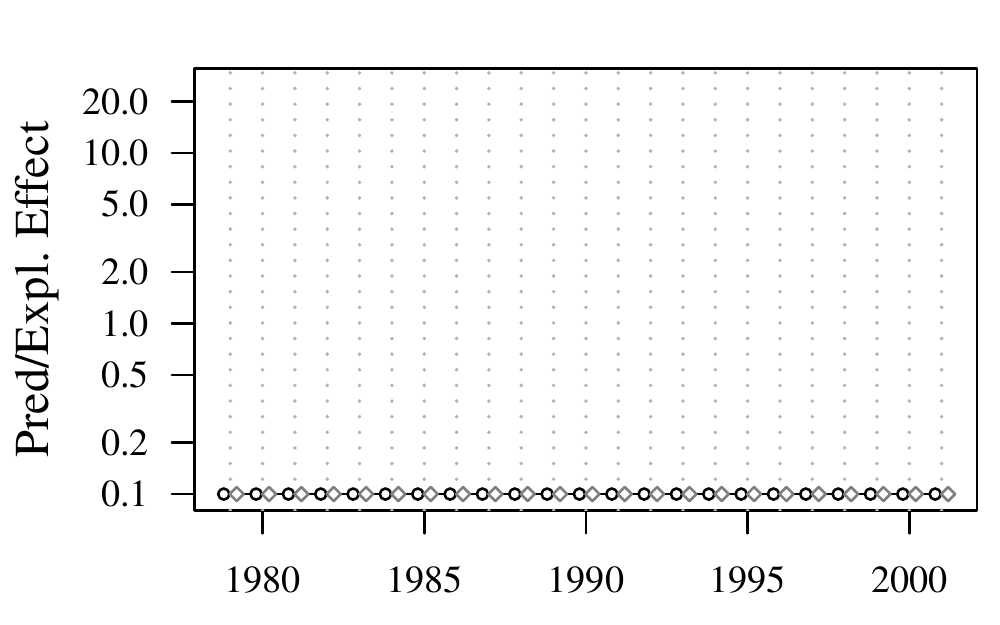} \\
\end{tabular} \\
Latent Space Distance \\
\includegraphics[width=7cm]{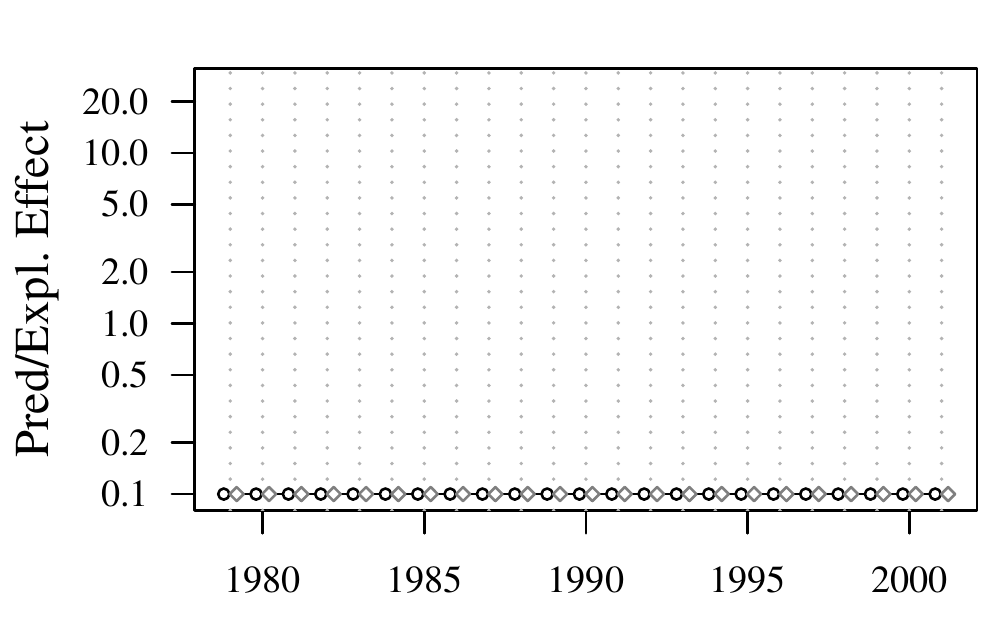} \\
\caption{Variable effects as measured by abs(elastic net $\beta$)/abs(logit $\beta$). When this quantity is  high, elastic net has not penalized the coefficient down and the variable can be said to be a stronger contributor to predictive performance. The ratios from the models based on a five-year lagged interval are in black, and those based on a ten-year lagged interval are in gray. To smooth the lines, we depict rolling means over a three-year period centered at the focal year. The points that are not shaded reflect years in which the coefficient ratio fell below 0.01, indicating that the variable was effectively removed from the model through regularization.} 
\label{releffects_dep}
\end{figure}

Our predictive tests reveal much about the roles played by each variable. 
We are able to consider how much a given variable contributes at different points in the time series under consideration. The plots in Figures \ref{releffects_covariates}--\ref{releffects_dep} give the abs(elastic net $\beta$)/abs(logit $\beta$). When this quantity is high, and above 1 especially, the corresponding variable has been selected and weighted highly as being important in predicting conflict. When is quantity is low, the corresponding variable has either been penalized completely out of the model or has been down weighted due to the variable's low contribution to the predictive performance of the model. Considering the ratio of the elastic net coefficient to the logistic regression coefficient provides a view of the degree to which the variable is critical to contributing to the prediction of conflict, given the simultaneous contributions of the rest of the variables. We note here that these feature-level summaries are intended to shed light on each feature's relative predictive contribution to the model (i.e., how much the magnitude of the variable's effect is deflated or elevated once the algorithm is designed to push effects towards zero when the variable does not contribute to predictive performance). This is different than characterizing the sign or shape of the relationship between the features and the dependent variable (i.e., marginal effects). For a flexible approach to characterizing marginal effects in complex statistical models, we refer readers to the partial derivative methodology proposed by \cite{Beck:2000}.  

Considering now the variable effects presented in Figures \ref{releffects_covariates}--\ref{releffects_dep}, we see several important broad themes when considered in the context of the established literature on international conflict. We consider the variables that perform well to be those that are consistently selected by the regularization procedure (i.e., are shaded points in the plots), and exhibit coefficient ratios near or above one. The variables that perform well include Flow (i.e., the product of sender initiations previously sent and recipient initiations previously received), memory, common community membership, contiguity, and shared IGO membership. Notably, most of the best performing predictive features are dependence effects, not exogenous covariates. Additionally, many of the covariates common in the conflict literalure are either rarely or never selected by the regularization method. Exogenous variables that are regularly kicked out of the model include trade dependence, defensive alliances, major power dyad, CINC ratio, and joint democracy. 

These patterns raise important implications for understanding and predicting international conflict. For nearly three decades, the quantitative study of conflict has been focused almost exclusively on the problem of predicting conflict (or the let lack thereof) on the basis of state and/or state-dyad attributes (i.e., exogenous covariates). Our results show that the dynamics of the interweaving system of dyadic conflicts may be just as important, if not more, in understanding the initiation of conflict. At the very least, our results serve as a call to scholars of international conflict to develop a theoreticaly informed model of conflict system dynamics with which to compare and/or integrate conventional covariate-based explanatory models.

\section{Conclusion}

We argue that predictive analysis, though it is statistically distinct from explanatory analyses, is a valuable tool for building explanatory models. We have shown that predictive analyses can be used to set benchmarks: to measure how much we know about an outcome, and to measure the improvement that a new analysis offers over its predecessors. We have further shown how predictive analysis can lead to insights, such as the length of the memory process involved in international conflict, that we can use it to understand the individual contributions of variables of interest, and that statistical significance does not necessarily imply that a variable is an important predictor. 

Our predictive exercise yields several interesting and compelling results. Ultimately, these results suggest that conflict is a long-memory process, that the simplest predictive algorithm, elastic net, is the most effective, that models with exogenous covariates alone generally perform worse than models based solely on the outcome variable, and that combined network-covariate models often do not provide a substantial improvement in predictive ability over the outcome-only benchmark model. Lastly, we see that several variables that are well established in the conflict literature contribute little to the prediction of conflict.  

We propose that predictive modeling is a promising means by which to enhance the study of political processes, particularly, though not exclusively, those for which we are unable to conduct controlled experiments or even use causal tools for observational data. In international politics for instance, one cannot experiment on conflict processes and the interconnectedness of states in the system precludes the use of matching techniques for causal inference (which require strict independence assumptions to produce valid estimates). But international relations does not stand alone with this problem, such situations occur frequently in American and comparative politics as well. In such cases especially, predictive modeling is a big and powerful tool that is too often left in the box.

\clearpage
\bibliographystyle{apsr}
\bibliography{BPWM}
\clearpage





\end{document}